\documentclass[twocolumn,showpacs,aps,longbibliography,prd]{revtex4-1}

\usepackage{epsf}
\usepackage{amsmath}
\usepackage{amsfonts}
\usepackage{amssymb}
\usepackage{bm}
\usepackage{bbm}
\usepackage{nicefrac}
\usepackage{xcolor}
\usepackage{pifont}
\usepackage{graphicx}
\usepackage{enumerate}
\usepackage{dcolumn}
\usepackage{maybemath}
\usepackage{xfrac}
\usepackage{siunitx}

\newcommand{\stdrule}{\rule[-1mm]{0mm}{4mm}}

\newcommand{\ee}{\mathrm{e}}
\newcommand{\ii}{\mathrm{i}}
\newcommand{\dd}{\mathrm{d}}

\newcommand{\calO}{\mathcal{O}}

\def\CaF2{{CaF$_2$}}

\definecolor{garrosgreen}{rgb}{0.1, 0.4, 0.1}
\definecolor{dartmouthgreen}{rgb}{0.05, 0.5, 0.06}
\definecolor{duelferred}{rgb}{0.7, 0.2, 0.1}
\definecolor{cambridgeblue}{rgb}{0.1, 0.3, 1.0}
\definecolor{oxfordblue}{rgb}{0.05, 0.2, 0.7}

\newcommand{\plus}{{\mbox{{\bf{\tiny +}}}}}
 
\def\calF{{\mathcal{F}}}
\def\calH{{\mathcal{H}}}
\def\calI{{\mathcal{I}}}
\def\calL{{\mathcal{L}}}

\def\calS{{\mathcal{S}}}
\def\calO{{\mathcal{O}}}
\def\calT{{\mathcal{T}}}

\begin{document}

\title{Gravitational Effects in $\bm g$ Factor Measurements and 
High--Precision Spectroscopy:\\
Limits of Einstein's Equivalence Principle}

\author{U. D. Jentschura}

\affiliation{Department of Physics,
Missouri University of Science and Technology,
Rolla, Missouri 65409, USA}

\affiliation{MTA--DE Particle Physics Research Group,
P.O.Box 51, H--4001 Debrecen, Hungary}

\affiliation{MTA Atomki, P.O.Box 51, H--4001 Debrecen, Hungary} 

\begin{abstract}
We study the interplay of general relativity, the equivalence
principle, and high-precision experiments involving
atomic transitions and $g$ factor measurements.
In particular, we derive a generalized Dirac Hamiltonian,
which describes both the gravitational coupling for
weak fields, as well as the electromagnetic coupling,
e.g., to a central Coulomb field.
An approximate form of this Hamiltonian is used
to derive the leading gravitational corrections to 
transition frequencies and $g$ factors. The position-dependence
of atomic transitions is shown to be compatible with the equivalence
principle, up to a very good approximation.
The compatibility of $g$ factor measurements 
requires a deeper, subtle analysis, in order to eventually
restore the compliance of $g$ factor measurements
with the equivalence principle.
Finally, we analyze small, but important limitations of
Einstein's equivalence principle due to
quantum effects, within high-precision experiments.
We also study the relation of these effects
to a conceivable gravitationally induced
collapse of a quantum mechanical wave function (Penrose conjecture),
and space-time noncommutativity,
and find that the competing effects should not preclude the 
measurability of the higher-order gravitational corrections.
In the course of the
discussion, a renormalized form of the Penrose conjecture is
proposed and confronted with experiment.
Surprisingly large higher-order gravitational effects are 
obtained for transitions in diatomic molecules.
\end{abstract}


\maketitle

\small

\tableofcontents

\normalsize

%
%
\section{Introduction}

According to Einstein's theory of gravitation,
space-time is locally flat, and the Einstein form of the
equivalence principle states that the
outcome of any non-gravitational
experiment should be independent of where
and when in the Universe it is performed.
Among the most accurately measured
quantities in physics, we find transition frequencies
in simple atomic systems and $g$ factor experiments,
both for free and bound leptons (electrons and muons).
Leptons are described, in curved
space-time, by the gravitationally and
electromagnetically coupled Dirac equation.
Here, we derive a generalized Dirac Hamiltonian
which describes both mentioned couplings for light fermions,
in electromagnetic and (weak) gravitational fields,
and establish its properties under a
particle-antiparticle transformation.
We also find its nonrelativistic form by
a Foldy--Wouthuysen transformation.

As convincingly demonstrated by the 
Shapiro delay, measured to
excellent accuracy by the Cassini spacecraft
in superior conjunction~\cite{BeIeTo2003}, we must assign a
coordinate dependence to the vacuum permittivity
and vacuum permeability, in global coordinates.
Based on these assumptions, we investigate the position-dependence
of  atomic transitions and
find (in agreement with Ref.~\cite{Wi1974prd})
that the position-dependence of their frequencies
is largely compatible with the equivalence principle.

For free and bound $g$ factor experiments,
gravitational frequency shifts of spin-flip transitions
have been the subject of rather intense 
discussions~\cite{MoFuSh2018ptep,MoFuSh2018remark1,MoFuSh2018remark2,Vi2018,Ni2018,Gu2018}.
Our paper addresses part of these questions but otherwise has
a much broader scope. Furthermore, it has long been conjectured that
subtle limitations to the Einstein equivalence
principle should occur within a full quantum theory.
We find such limitations, both due to the Fokker 
precession as well as due to the noncommutativity of the electron's
momentum operator with the global space-time coordinates.

It is our goal to present a comprehensive and 
relatively 
easily digestible account of related matters,
despite the length of the current article.
For clarification, we should point out that 
throughout this paper, we consider gravitational effects for an 
atom at rest with respect to a center of gravity, in contrast to 
Refs.~\cite{Pa1980prd,PaPi1982,AdChVa2012},
where the authors refer to an atom in a 
freely falling reference frame.
Note that in Ref.~\cite{AuMa1994},
the results of Refs.~\cite{Pa1980prd,PaPi1982,AdChVa2012} are
generalized to accelerated and rotating 
reference frames; such frames are not of interest for the 
current study.
Furthermore, we assume, throughout the 
paper, that local Lorentz invariance
is conserved. Conceivable correction terms beyond
this approximation are considered 
in Ref.~\cite{GaHa1990}.
In detail, the paper is organized as follows.
In Sec.~\ref{sec2}, we consider the 
gravitationally and electromagnetically coupled
Dirac equation, and the scaling of atomic transition
frequencies, and bound-state $g$ factors,
induced by the gravitational coupling.
The interrelation of quantum mechanics and 
Einstein's equivalence principle is 
being studied in Sec.~\ref{sec3}.
Roughly speaking, the question is whether a 
non-deterministic theory (namely, quantum mechanics)
can in principle be fully compatible with a
fully deterministic theory (namely, general relativity),
given the fact that position and momentum operators 
in quantum mechanics behave differently from their
classical counterparts.
We shall find tiny, but important corrections
to the so-called $\sqrt{T}$ scaling which otherwise 
ensures the compatibility of the gravitationally
corrected frequencies with the equivalence principle.
The measurability of the higher-order gravitational 
corrections is discussed in Sec.~\ref{sec4}.
Conclusions are reserved for Sec.~\ref{sec5}.

%
%
\section{Gravity and Scaling}
\label{sec2}

%
%
\subsection{Coupled Dirac Hamiltonian}
\label{sec2A}

We use units with $\hbar = c = \epsilon_0 = 1$.
The relativistic (gravitationally coupled)
Dirac--Schwarzschild Hamiltonian
is~\cite{JeNo2013pra}
\begin{equation}
\label{HDS}
H_{\rm DS} =
\frac12 \, \left\{ \vec\alpha \cdot \vec p,
\left( 1 - \frac{r_s}{r} \right) \right\} +
\beta m \left( 1 - \frac{r_s}{2 r} \right) \,,
\end{equation}
where
$\vec \alpha =
\left( \begin{array}{cc}
0 & \vec \sigma \\
\vec \sigma & 0
\end{array} \right)$
and 
$\beta =
\left( \begin{array}{cc}
\mathbbm{1}_{2 \times 2}  & 0 \\
0 & \mathbbm{1}_{2 \times 2}
\end{array} \right)$ are Dirac matrices,
and $m$ denotes the fermion (electron) mass.
The $(2 \times 2)$ Pauli
spin matrices are denoted as $\vec \sigma$.
Hermiticity properties of 
this Hamiltonian are discussed in App.~\ref{appA1},
while a comparison of this result to 
other literature references is the subject
of App.~\ref{appA2}.

The Foldy--Wouthuysen (FW) transformed 
Dirac--Schwarzschild Hamiltonian 
has been found in Ref.~\cite{JeNo2013pra},
\begin{subequations}
\label{HDSFW}
\begin{align}
H_{\rm FW} =& \;
\beta \, \left( m + \frac{\vec p^{\,2}}{2 m} -
\frac{\vec p^{\,4}}{8 m^3} 
- \beta \, \frac{m \, r_s}{2 \, r} 
- \frac{3 r_s}{8 m} \,
\left\{ \vec p^{\,2}, \frac{1}{r} \right\} \right.
\nonumber\\
& \; + \left.
\frac{3 \pi r_s}{4 m} \delta^{(3)}(\vec r) \,
+ \frac{3 r_s}{8 m} \, \frac{\vec\Sigma \cdot \vec L}{r^3} \right) \,,
\end{align}
which can be reformulated as
\begin{multline}
H_{\rm FW} = \beta \, 
\left( m + \frac{\vec p^{\,2}}{2 m} \right) 
\\
- \beta \, \left( \frac{\vec p^{\,4}}{8 m^3} - \frac{m \, r_s}{2 \, r}
- \frac{3}{16 m} \left\{ \vec \Sigma \cdot \vec p,
\left\{ \vec \Sigma \cdot \vec p, \frac{r_s}{r} \right\} \right\}
\right),
\end{multline}
\end{subequations}
where $r_s$ is the Schwarzschild radius, and
$r$ is the radial variable in Eddington coordinates~\cite{Ed1924}.
The latter form is obtained from the first 
by applying the operator identity
$\{ A, \{ A, B \} \} =
2 \{ A^2, B \} - [ A, [A, B]] $,
for $A = \vec \Sigma \cdot \vec p$ and $B = 1/r$,
where the $(4 \times 4)$ spin matrices are
$\vec \Sigma = \left( \begin{array}{cc}
\vec\sigma & 0 \\
0 & \vec\sigma \end{array} \right)$.
The generalization of the Dirac--Schwarzschild
Hamiltonian~\eqref{HDS} to the case of an additional
external electromagnetic fields
[denoted here as the Dirac--Schwarzschild--Coulomb
(DSC) Hamiltonian]
involves the replacement of the
kinetic momentum operators $\vec p$ by the
canonical momentum operators $\vec \pi = \vec p - e \, \vec A$,
and the addition of the scalar potential term $e \, A^0$.
Here, $e = -|e|$ is the
physical electron charge.
It reads as follows,
\begin{equation}
\label{HDSC}
H_{\rm DSC} =
\frac12 \, \left\{ \left( 1 - \frac{r_s}{r} \right),
\vec\alpha \cdot \vec \pi \right\} +
e \, A^0 \mathbbm{1}_{4 \times 4} + 
\beta m \left( 1 - \frac{r_s}{2 r} \right).
\end{equation}
After a Foldy--Wouthuysen 
transformation, one obtains the Hamiltonian $H_{\rm EM}$
which describes the coupling to 
external electromagnetic fields,
\begin{multline}
\label{HEM}
H_{\rm EM} =
\beta \, \left( m + 
\frac{(\vec\Sigma \cdot \vec\pi)^2}{2 m} -
\frac{(\vec\Sigma \cdot \vec\pi)^4}{8 m^3} \right)
+ e \, A^0 \mathbbm{1}_{2 \times 2}
\\[1.0ex]
- \beta \, \left( \frac{m \, r_s}{2 \, r}
+ \frac{3}{16 m} 
\left\{ \vec \Sigma \cdot \vec \pi,
\left\{ \vec \Sigma \cdot \vec \pi, 
\frac{r_s}{r} \right\} \right\} \right)
\\[1.0ex]
+ \left\{ 1 + \frac{r_s}{r}, \frac{e}{16 m^2} \,
\left( \vec\nabla \cdot \vec E +
\vec \Sigma \cdot (\vec E \times \vec \pi -
\vec \pi \times \vec E)
\right) \right\} \,.
\end{multline}
This Hamiltonian is a $(4 \times 4)$-matrix,
diagonal in the space of $(2 \times 2)$-submatrices.
The $(2 \times 2)$-particle Hamiltonian 
$ H^+_{\rm EM}$ is obtained by 
replacing $\beta \to 1$:
\begin{multline}
\label{HEMplus}
H^+_{\rm EM} = m +
\frac{(\vec\sigma \cdot \vec\pi)^2}{2 m} -
\frac{(\vec \sigma \cdot \vec\pi)^{4}}{8 m^3} 
+ e \, A^0
\\[0.1133ex]
- \frac{m \, r_s}{2 \, r}
- \frac{3}{16 m}
\left\{ \vec \sigma \cdot \vec \pi,
\left\{ \vec \sigma \cdot \vec \pi,
\frac{r_s}{r} \right\} \right\} 
\\[0.1133ex]
+ \left\{ 1 + \frac{r_s}{r}, \frac{e}{16 m^2} \,
\left( \vec\nabla \cdot \vec E +
\vec \sigma \cdot (\vec E \times \vec \pi -
\vec \pi \times \vec E)
\right) \right\} \,.
\end{multline}
The antiparticle Hamiltonian $H^-_{\rm EM}$ 
is obtained from $H_{\rm EM}$ by 
replacing $\beta \to -1$, taking into account an overall factor $-1$
due to the reinterpretation principle,
replacing $\vec \Sigma \to -\vec\sigma$, 
and $\vec p \to - \vec p$, again due to 
reinterpretation for antiparticles.
One can convince oneself that the antiparticle Hamiltonian 
$H^-_{\rm EM}$ can be obtained from the 
particle Hamiltonian $H^+_{\rm EM}$ by the replacement $e \to -e$
(charge conjugation, hence $\vec \pi \to \vec \pi' = \vec p + e \, \vec A$)), 
while all the gravitational terms
are invariant under the particle-antiparticle transformations~\cite{JeNo2013pra},
establishing the equivalence principle for anti-particles.

We now continue to work with the particle Hamiltonian~\eqref{HEMplus},
which can be simplified based on the identity
$(\vec\sigma \cdot \vec \pi)^2 
= \vec \pi^2 - e \, \vec \sigma \cdot \vec B$, which implies that
\begin{multline}
\label{HFW}
H^+_{\rm EM} = m + \frac{\vec \pi^{\,2}}{2 m}
- \frac{\vec \pi^{\,4}}{8 m^3} 
- \frac{e}{2 m} \, \vec \sigma \cdot \vec B 
\\[0.1133ex]
+ e \, A^0
+ \frac{e}{8 m^3} \, \{ \vec \sigma \cdot \vec B, \vec\pi^{\,2} \}
- \frac{m \, r_s}{2 \, r}
+ \frac{3 \pi r_s}{4 m} \delta^{(3)}(\vec r)
\\[0.1133ex]
- \frac{3}{8 m}
\left\{ \vec \pi^{\,2} - e \, \vec \sigma \cdot \vec B,
\frac{r_s}{r} \right\}
+ \frac{3 r_s}{8 m r^3}
\vec \sigma \cdot \vec r \times \vec \pi 
\\[0.1133ex]
+ \left\{ 1 + \frac{r_s}{r}, \frac{e}{16 m^2} \,
\left( \vec\nabla \cdot \vec E +
\vec \sigma \cdot \vec E \times \vec \pi -
\vec \sigma \cdot \vec \pi \times \vec E
\right) \right\} \,.
\end{multline}
We should note that related calculations have 
recently been considered in 
other contexts~\cite{ObSiTe2014,ObSiTe2016,ObSiTe2017},
with an important clarifying remark given in the text following
Eq.~(7.33) of Ref.~\cite{ObSiTe2017}
(see also Ref.~\cite{JeNo2014jpa}).

We now discuss a general metric for weak gravitational fields and 
gravitational red shifts. For inspiration,
we start with the Schwarzschild metric~\cite{Sc1916}
in isotropic form (Sec.~43 of Chap.~3 of Ref.~\cite{Ed1924}),
\begin{subequations}
\begin{equation}
\label{eddington}
\dd s^2 =
\left( \frac{1 - r_s/(4 r)}{1 + r_s/(4 r)} \right)^2 \, \dd t^2
-\left( 1 + \frac{r_s}{4 r} \right)^4 \; \dd \vec r^{\,2} \,.
\end{equation}
This metric can be expanded to first order in the 
potential $\Phi(\vec r) = - G M/r$ ,
where $M$ is the mass of the central gravitational object,
and generalized 
to arbitrary (weak) gravitational potentials $\Phi$,
\begin{align}
\label{ds2}
\dd s^2 
= & \; \left( 1 - \frac{r_s}{r} \right) \, \dd t^2
- \left( 1 + \frac{r_s}{r} \right) \, \dd \vec r^{\,2}
\nonumber\\[0.1133ex]
=& \; \left( 1 + 2 \, \Phi \right) \, \dd t^2
- \left( 1 - 2 \, \Phi \right) \, \dd \vec r^{\,2}
\nonumber\\[0.1133ex]
=& \; T \, \dd t^2 - H \, \dd \vec r^{\,2}
= \overline g_{\mu\nu} \, \dd x^\mu \, \dd x^\nu \,.
\end{align}
\end{subequations}
Here, $\overline g_{\mu\nu} = {\rm diag}(T, -H, -H,-H)$ is the 
curved-space metric, while we reserve the symbol
$\widetilde g_{\mu\nu} $ for the metric of free space~\cite{JeNo2013pra}.
In the following, 
we use the symbols $T$ and $H$ for the case of a general 
gravitational potential $\Phi$.

In a metric of the form~\eqref{ds2}
(see Refs.~\cite{OhRu1994,As2002,Pa2010})
one has for light, which travels on a zero geodesic
with $\dd s^2 = 0$, 
\begin{equation}
\label{shapiro}
\left| \frac{\dd \vec r }{ \dd t } \right|^2 = 
\frac{ 1 + 2 \, \Phi }{ 1 - 2 \, \Phi } = 
\frac{T}{H} \approx 1 + 4 \, \Phi \,.
\end{equation}
We thus generalize~\eqref{shapiro} 
to general gravitational fields.
The Shapiro time delay~\cite{Sh1964,ShEtAl1968,Sh1999,Lo1988,KrTr1988}
 is consistent
with an effective speed of light, of the form
$c_{\rm eff} = 1 + 2 \, \Phi = \sqrt{ T/H } $,
to first order in the gravitational potential.
This implies that in electrodynamics, 
we must assign a slight gravitational dependence
to the vacuum permittivity $\epsilon$ and vacuum permeability $\mu$,
so that 
\begin{equation}
\label{ceff}
c_{\rm eff}^2 = \frac{1}{\epsilon \, \mu} = \frac{T}{H} \,,
\qquad
\epsilon = \mu = \sqrt{ \frac{H}{T} } \,,
\end{equation}
consistent with Eq.~(4) of Ref.~\cite{Wi1974prd}.

%
%
\subsection{Gravity and atomic transitions}
\label{sec2B}

The generalization of the Hamiltonian~\eqref{HDSC} 
to a general gravitational potential $\Phi$
can be found by realizing that the derivation,
outlined in Ref.~\cite{JeNo2013pra}, goes through 
for a general metric of the form given in Eq.~\eqref{ds2}.
The Hamiltonian reads as
\begin{align}
H_{\rm DSC} =& \;
\frac12 \, \left\{ 1 + 2 \Phi,
\vec\alpha \cdot \vec \pi \right\} +
e \, A^0 + \beta m \left( 1 + \Phi \right) 
\\[0.1133ex]
=& \; \frac12 \, \left\{ \sqrt{\frac{T}{H}},
\vec\alpha \cdot \vec \pi \right\} +
e \, A^0 + \beta m \sqrt{T} \,.
\end{align}
If we ignore commutators of the gravitational fields
and the momentum operators, then
we may approximate
\begin{equation}
\label{HDSCapprox}
H_{\rm DSC} \approx
\sqrt{\frac{T}{H}} \, \vec \alpha \cdot \vec \pi
+ \sqrt{T} \, \beta \, m
+ e \, A^0 \,.
\end{equation}
We here confirm the result given in Eq.~(14) of Ref.~\cite{Wi1974prd},
and show that anticommutators are needed in 
order to  turn the Hamiltonian into a manifestly
Hermitian entity.
The approximation~\eqref{HDSCapprox} is valid 
if we assume that $T$ and $H$ remain constant to very 
good approximation, over the distance scales
relevant to the described quantum mechanical phenomena.

We consider the Hamiltonian~\eqref{HDSCapprox}
for the case $\vec A = \vec 0$, and 
$e \, A^0 = - Z e^2/(4 \pi \epsilon |\vec \rho|)$,
where $|\vec \rho|$ is the distance to the atomic nucleus.
In this case, the Hamiltonian becomes
\begin{equation}
\label{HDSCproblem}
H_{\rm DSC} = \sqrt{\frac{T}{H}} \, \vec \alpha \cdot \vec p 
+ \sqrt{T} \, \beta \, m
- \frac{Z e^2}{4 \pi \epsilon |\vec \rho|} \,,
\end{equation}
where the subscripts refer to Dirac, Schwarzschild and Coulomb (DSC).
The energy eigenvalue equation is
\begin{equation}
H_{\rm DSC} \, \psi = E \, \psi \,.
\end{equation}
With Ref.~\cite{Wi1974prd}, we now perform the following
scaling,
\begin{equation}
\label{scaling}
m = \bar m \, \frac{1}{\sqrt{H}}  \,, \qquad
e^2 = \bar e^2 \sqrt{\frac{T}{H}} \, \epsilon \,, \qquad
E = \bar E \, \sqrt{\frac{T}{H}} \,,
\end{equation}
which turns the eigenvalue problem~\eqref{HDSCproblem} into
\begin{equation}
\label{saye1}
\left( \vec \alpha \cdot \vec p
+ \beta \, \bar m
- \frac{Z \bar e^2}{4 \pi |\vec \rho|} \right) \, \psi =
\bar E \, \psi \,.
\end{equation}
The energy can be given in terms of the scaled function 
$f(n, J, Z\alpha)$ which has been introduced by 
Sapirstein and Yennie in Ref.~\cite{SaYe1990},
\begin{subequations}
\label{escaling}
\begin{align}
\label{saye2}
\bar E =& \; \bar m \, f(n,J, Z \bar \alpha) \,,
\\[0.1133ex]
f(n,J, Z \bar \alpha) =& \; \left( 1 + \frac{ (Z \bar \alpha)^2 }%
{ [n_r + \sqrt{ (J + 1/2)^2 - (Z\bar\alpha)^2 } ]^2 } \right)^{-\tfrac12} \,,
\end{align}
where $n_r = n - J - 1/2$ is the 
so-called reduced principal quantum number.
The electron's orbital angular momentum quantum number is 
$\ell$, while its total angular momentum is $J$.
Finally, the gravitationally ``modified'' (as it turns out, invariant) 
fine-structure constant is 
\begin{equation}
\label{alphaconst}
\bar \alpha 
= \frac{\bar e^2}{4 \pi} 
= \sqrt{\frac{H}{T}} \, \frac{e^2}{4 \pi \epsilon} 
= \sqrt{\frac{H}{T}} \, \frac{e^2}{4 \pi} \, \sqrt{\frac{T}{H}} 
= \frac{e^2}{4 \pi} = \alpha \,.
\end{equation}
\end{subequations}
The position-independence of the fine-structure 
constant has been verified experimentally,
in a dedicated experiment described in 
Ref.~\cite{TuEtAl1983}.
We should notice that experimental possibilities
to search for a temporal as well as 
spatial variation of the fine-structure 
constant have since dramatically improved
in accuracy~\cite{FiEtAl2004,PeEtAl2004,GoEtAl2014,HuEtAl2014}.
The scaling of the bound-state energy
is found as
\begin{equation}
\label{yes}
E = \sqrt{\frac{T}{H}} \, \bar E 
= \sqrt{\frac{T}{H}} \, \bar m \,  f(n,J, Z \alpha) 
= \sqrt{T} \, m \,  f(n,J, Z \alpha) \,,
\end{equation}
valid for both main-structure (change in the principal 
quantum number) as well as fine-structure transitions.

%
%
\subsection{Gravity and $\maybebm{g}$ factor}
\label{sec2C}

We start from Eq.~\eqref{HDSCapprox},
but this time we include the static vector potential 
$\vec A = \frac12 \, (\vec B \times \vec r)$,
which describes a constant $\vec B$ field.  Hence,
$H_{\rm DSC}$ attains the form
\begin{equation}
\label{HDSCapprox_full}
H_{\rm DSC} = \sqrt{\frac{T}{H}} \, \vec \alpha \cdot 
(\vec p  - e \, \vec A)
+ \sqrt{T} \, \beta \, m + e \, A^0 \,.
\end{equation}
Taking into account that $\vec A = \tfrac12 \, (\vec B \times \vec r)$,
one can write the Hamiltonian $H_M$ which describes the 
magnetic coupling of the electron to the 
external field as 
\begin{equation}
\label{HM}
H_M = - \sqrt{\frac{T}{H}} \, e \,
\, \vec \alpha \cdot \vec A 
= - \sqrt{\frac{T}{H}} \, \frac{e}{2} \,
\, \vec \alpha \cdot (\vec B \times \vec r) \,.
\end{equation}
Canonically, one assumes that $\vec B$ is directed
along the $z$ axis~\cite{GlSh2002}.
The Land\'{e} $g$ factor 
(written as $g_J$) and the expectation value for a hydrogenic 
state in a homogeneous $\vec B$ field can be 
expressed as 
\begin{equation}
\label{HMexpec}
\left< H_M \right> = 
- g_J \, \frac{e}{2 m}  \, | \vec B | \, \mu_J \,,
\end{equation}
where $\mu_J$ is the projection of the 
electron's total angular momentum 
(angular$+$spin) onto the axis of the $\vec B$ field.
The expectation value in Eq.~\eqref{HMexpec}
is to be taken in an eigenstate of the unperturbed
problem, i.e., in a (gravitationally modified)
Dirac--Coulomb eigenstate of the Hamiltonian~\eqref{HDSCproblem}.

Let us therefore consider the Hamiltonian~\eqref{HDSCapprox_full}
under the same scaling as the one used in Eq.~\eqref{scaling}.
The eigenvalue problem transforms into
\begin{equation}
\left( \vec \alpha \cdot (\vec p - \bar e\, \vec A)
+ \beta \, \bar m
- \frac{Z \bar e^2}{4 \pi |\vec \rho|} \right) \, \psi =
\bar E \, \psi \,.
\end{equation}
For the magnetic-field coupling Hamiltonian,
written in terms of the scaled variables,
\begin{equation}
\label{HMbar}
\bar H_M = - \frac{\bar e}{2} \,
\, \vec \alpha \cdot (\vec B \times \vec r) \,,
\end{equation}
we therefore have the following relation,
which holds  in view of the analogy with the 
unperturbed Dirac problem (see Ref.~\cite{GlSh2002}):
\begin{align} 
\left< \bar H_M \right> =& \;
- \bar g_J \, \frac{\bar e}{2 \bar m}  \, | \vec B | \, \mu_J \,,
\\[0.1133ex]
\bar g_J =& \; \frac{\varkappa}{J (J+1)} \left( \varkappa \,
\frac{\bar E}{\bar m} - \frac12 \right) 
\nonumber\\[0.1133ex]
=& \; \frac{\varkappa}{J (J+1)} \left( \varkappa \,
f(n,J, Z \alpha) - \frac12 \right) \,.
\end{align} 
Here, $\varkappa$ is the Dirac angular quantum 
number, which is given as 
$\varkappa = (-1)^{J+ \ell +1/2} \, \left(J + \frac12 \right) $.
A comparison of the Hamiltonian $H_M$ given in Eq.~\eqref{HM}
to the Hamiltonian $\bar H_M$ given in Eq.~\eqref{HMbar} 
reveals that  $\left< H_M \right> =
-(T/H)^{1/2} \, \left< \bar H_M \right> $
so that, in view of Eq.~\eqref{HMexpec},
\begin{equation}
\label{gJTH}
g_J = \frac{\sqrt{T}}{H} \, \bar g_J \, 
= \frac{\sqrt{T}}{H} \, 
\frac{\varkappa}{J (J+1)} \left( \varkappa \,
f(n,J, Z \alpha) - \frac12 \right) \,.
\end{equation}
For the ground state, one has with $\varkappa = -1$
and $J = 1/2$,
\begin{equation}
\label{bargJTH}
g_J = \frac{\sqrt{T}}{H} \, 
\frac43 \, \left( \sqrt{1 - (Z\alpha)^2} + \frac12 \right) \,.
\end{equation}
The free-electron $g$ factor is obtained from this 
expression, in the limit $Z\alpha \to 0$, and is equal to 
$g_S = 2 \sqrt{T}/H$. One can convince oneself that this result is
compatible with the terms proportional to $\vec \sigma \cdot \vec B$
in Eq.~\eqref{HFW}; these determine the g factor.

At this stage, we have clarified the gravitational corrections
to the {\em non-anomalous} part of the electron's magnetic moment.
For the anomalous part, we need to consider the
generalized Dirac equation, which necessitates
the introduction of form factors.
We recall that in flat space, 
the electromagnetically coupled 
Dirac equation reads as 
$\left[ \widetilde\gamma^\mu (p_\mu - e \, A_\mu) - m \right] \, \psi = 0$,
where the $\widetilde\gamma^\mu$ are Dirac $\gamma$ 
matrices which fulfill the anti-commutator relations
$\{ \widetilde\gamma^\mu, \widetilde\gamma^\nu \}
= 2 \widetilde g^{\mu\nu} 
= {\rm diag}(1, -1, -1, -1)$. In order to describe the 
anomalous magnetic moment,
one replaces the Dirac $\gamma$ matrices by
a form-factor expression (see Chap.~7 of Ref.~\cite{ItZu1980}),
\begin{equation}
\label{formfactor}
\widetilde\gamma^\mu \to \widetilde\gamma^\mu \, F_1(q^2) + 
\frac{\ii \widetilde\sigma^{\mu\nu} \, q_\nu}{2 m} F_2(q^2) \,,
\end{equation}
where the spin matrices are given as $\widetilde\sigma^{\mu\nu} = 
\tfrac{\ii}{2} [ \widetilde\gamma^\mu, \widetilde\gamma^\nu ]$.
The replacement leads to the modified Dirac 
(MD) Hamiltonian~\cite{JePa2002},
\begin{align}
\label{HDDM}
H_{\rm MD} =& \; \vec{\alpha} \cdot
\left[\vec{p} - e \, F_1(\vec\nabla^2 ) \, \vec{A}\right]
     + \beta\,m + {\mathrm e} \, F_1(\vec\nabla^2 ) \, A^0 
\nonumber\\[0.1133ex]
& \; + F_2(\vec\nabla^2 ) \, \frac{e}{2\,m} \, \left({\mathrm i}\,
\vec{\gamma} \cdot \vec{E} - \beta \, \vec{\sigma} \cdot \vec{B} \right) \,.
\end{align}
In the following, we shall approximate
\begin{equation}
\label{approx}
F_1(q^2) \approx F_1(0) = 1 \,,
\qquad
F_2(q^2) \approx F_2(0) = \kappa \approx \frac{\alpha}{2 \pi} \,,
\end{equation}
and set the external electric field equal to zero, $\vec E = \vec 0$.
[We remember that, if we set $\vec E$ equal to the 
Coulomb electric field, the corresponding term in Eq.~\eqref{HDDM}
describes the anomalous magnetic-moment correction to the Lamb shift.]
Here, $\kappa$ describes the anomalous magnetic moment 
correction to the electron's spin $g$ factor,
and is approximated by the Schwinger term $\alpha/(2 \pi)$.

With the approximations outlined in Eq.~\eqref{approx},
the Hamiltonian~\eqref{HDDM} becomes
\begin{equation}
\label{HDm}
H_{\rm MD} = \vec{\alpha} \cdot
\left(\vec{p} - e \, \vec{A}\right)
+ \beta\,m + e \,  A^0 - 
\kappa \, \frac{e}{2\,m} \,
\beta \, \vec{\sigma} \cdot \vec{B} \,.
\end{equation}
We carry out a  replacement analogous to Eq.~\eqref{formfactor} in curved space,
\begin{equation}
\overline\gamma^\mu \to \overline\gamma^\mu \, F_1(q^2) +
\frac{\ii \overline\sigma^{\mu\nu} \, q_\nu}{2 m} F_2(q^2) \,,
\qquad
\{ \overline\gamma^\mu, \overline\gamma^\nu \} = 
2\overline g^{\mu\nu} \,,
\end{equation}
where 
$\overline g^{\mu\nu} = {\rm diag}(1/T, -1/H, -1/H, -1/H)$
is the inverse of the metric $\overline g_{\mu\nu}$ given 
in Eq.~\eqref{ds2}. The curved-space 
Dirac spin matrices $\overline\sigma^{\mu\nu} = \frac12 \,
[ \overline\gamma^\mu, \overline\gamma^\nu ]$ fulfill
$\overline\sigma^{\mu\nu} = 
\widetilde\sigma^{\mu\nu} / H$,
and the gravitational modification of 
Eq.~\eqref{HDm} reads as
\begin{equation}
\label{HX}
H = \sqrt{\frac{T}{H}} \, \vec \alpha \cdot
(\vec p  - e \, \vec A)
+ \sqrt{T} \, \beta \, m
+ e \, A^0 
- \frac{\sqrt{T}}{H} \, \kappa \, \frac{e}{2\,m} \,
\beta \, \vec{\sigma} \cdot \vec{B} \,.
\end{equation}
The gravitationally modified 
electron $g_J$ factor (for the $1S$ state) thus is, in view of 
Eq.~\eqref{bargJTH} and Eq.~\eqref{HX},
\begin{equation}
\label{gJ1S}
g_J = \frac{\sqrt{T}}{H} \, 
\frac43 \, \left( \sqrt{1 - (Z\alpha)^2} + \frac12 + 
\frac32 \, \kappa \right) \,,
\end{equation}
where the free-electron term is obtained in the limit 
$Z\alpha \to 0$. The scaling with $\sqrt{T}/H$ is thus established as a 
universal scaling of the free-electron and bound-electron $g$
factors, including the anomalous-magnetic-moment 
correction. 

%
%
\subsection{Equivalence principle and $\maybebm{g}$ factor}
\label{sec2D}

According to Eq.~\eqref{yes}, atomic transition frequencies receive a 
gravitational correction proportional to $\sqrt{T}$, 
while according to Eq.~\eqref{gJ1S},
the $1S$ electron $g$ factor
receives a correction proportional to 
$\sqrt{T}/H$.
The prefactor $\sqrt{T}$ in Eq.~\eqref{yes} describes the 
transition from coordinate time to laboratory time, 
This is evident from the metric~\eqref{ds2},
$\dd s^2 = T \, \dd t^2 - H \, \dd \vec r^{\,2} = \dd \tau^2$,
where $ \dd \tau^2 $ measures the (square of the) time 
interval in the local Lorentz frame.
We can convert the time derivative operator from 
coordinate time to the time elapsed in the local Lorentz frame,
\begin{equation}
\sqrt{T} \, \dd t = \dd \tau \,,
\qquad
\ii \frac{\partial}{\partial \tau} = 
\frac{\ii}{\sqrt{T}} \, \frac{\partial}{\partial t} \,.
\end{equation}
The energy~\eqref{yes} is formulated with respect to the 
coordinate time, and so, the energy in the laboratory 
can be obtained by dividing the energy $E$ given in Eq.~\eqref{yes}
by a factor $1/\sqrt{T}$, and 
one obtains the laboratory atomic energy levels as being
given by the expression $ m \,  f(n,J, Z \alpha) $.

Let us put this statement into the context 
of the weak and strong forms of the equivalence principle.
The ``weak equivalence principle'' (WEP) asserts the 
proportionality of ``mass'' (``inertial mass'') and  ``weight'' (which enters
the gravitational force law).
The Einstein equivalence principle (EEP) states that {\em (i)} WEP is valid,
{\em (ii)} the outcome of any local non-gravitational experiment is independent
of the velocity of the freely-falling reference frame in which it is performed
(local Lorentz invariance, LLI), and
{\em (iii)} the outcome of any local non-gravitational experiment is
independent of where and when in the universe it is performed
(local position invariance, LPI).

The scaling with $\sqrt{T}/H$ of the electron's $g$ factor,
in coordinate time, taken at face value, would 
imply a scaling with $1/H$ in the local Lorentz frame
of each laboratory, after dividing out the factor $\sqrt{T}$.
This would make the outcome of a non-gravitational experiment
(the measurement of the electron's $g$ factor)
dependent on the position,
limit the validity of the principle of local position invariance, and, hence, the EEP.

In order to resolve the problem,
we note that we have assumed, in our derivation, that the 
$\vec A$ field is given in terms of 
the components of the covariant basis,
\begin{equation}
\vec A = A^i \, \vec e_i \,,
\qquad
\vec e_i \cdot \vec e_j = H \, \delta_{ij} \,.
\end{equation}
Latin indices indicate spatial components ($i,j,k,\ldots = 1,2,3$).
However, the ``Cartesian'' unit vectors (index $c$)
which span the local Lorentz frame are
\begin{equation}
\hat e_i = \frac{1}{\sqrt{H}} \, \vec e_i \,,
\qquad
\hat e_i \cdot \hat e_j = \delta_{ij} \,.
\end{equation}
Let $x^i$ denote the components of the 
position vector $\vec r$ in the basis spanned by the 
$\vec e_i $, while the components $x_c^j $ are relevant to the 
basis spanned by the $\hat e_i$.  Then,
\begin{equation}
x_c^j = \sqrt{H} \, x^j \,,
\qquad
A_c^i = \sqrt{H} \, A^j \,.
\end{equation}
We denote by $\epsilon^{ijk}$ the totally antisymmetric 
Levi-Civit\`{a} tensor (under the normalization 
$\epsilon^{123} =1 $). Then, we have
\begin{equation}
\vec A = \frac12 \, \vec B_c \times \vec r_c
= \frac12 \, \hat e_i \, \epsilon^{ijk} \, B^j \, x_c^k \,,
\end{equation}
which is the appropriate vector potential for a
magnetic field with ``Cartesian'' components $B_c^i$,
measured in the local Lorentz frame.
The curl of $\vec A$ enters into Eq.~\eqref{bargJTH};
it is calculated with 
momentum operators $\vec p = -\ii \vec\nabla$ where
$\nabla^k = -\ii \partial/\partial x^k$, and hence,
\begin{equation}
B^i = \epsilon^{ijk} \frac{\partial}{\partial x^j} A^k
= \frac12 \, \epsilon^{ijk} \frac{\partial}{\partial x^j}
\epsilon^{k \ell m} \, B_c^\ell \, \sqrt{H} \, x^m
= \sqrt{H} \, B_c^i  \,.
\end{equation}
For the vector $\vec B$, this means that
\begin{equation}
\label{ottifant}
\vec B = B^i \, \vec e_i
= (\sqrt{H}\, B_c^i) \, (\sqrt{H} \hat e_i )
= H\, B_c^i \, \hat e_i \,.
\end{equation}
Thus, the $i$th component of $\vec B$, written in our
basis, is equal to $H$ times the $B$ field measured by a local
observer, in his or her own Lorentz frame.
This implies that, when normalized to the local $B$ field,
spin-flip frequencies transform with a factor 
$\sqrt{T}$, not $\sqrt{T}/H$, respecting the equivalence principle.
We note that the same factor $H$ is obtained in Ref.~\cite{Wi1974prd}
for the transformation of the hyperfine-structure generating 
$B$ field of a nucleus, from global coordinates to the local
Lorentz frame; however, the derivation proceeds in a completely 
different way [see Eqs.~(32)---(34) of Ref.~\cite{Wi1974prd}].
One notes that the restoration of the $\sqrt{T}$ scaling 
actually is absolutely crucial for the validity of the current 
adjustment of the fundamental constants~\cite{MoNeTa2016}.

%
%
\section{Quantum Mechanics and Equivalence Principle}
\label{sec3}

%
%
\subsection{Leading order
and $\maybebm{\sqrt{T}}$ scaling}
\label{sec3A}

We recall, from Sec.~\ref{sec2A},
that relatively weak gravitational fields give rise
to a metric
\begin{align}
\dd s^2 =& \; T \, \dd t^2 - H \, \dd \vec r^{\,2} \,,
\\[0.1133ex]
T =& \; 1 + 2 \, \Phi \,, \qquad H = 1 - 2 \, \Phi \,,
\end{align}
where $\Phi$ is the gravitational potential.
Hence, if, in global coordinates, an energy goes as
\begin{equation}
\label{common_prefactor}
E = \sqrt{T} \, E_c \,,
\end{equation}
where $E_c$ is the energy measured in a 
local, Cartesian Lorentz frame, then this effect is 
physically unobservable if the experiment 
is carried out locally, because 
the time derivative operator $\dd/\dd \tau$ 
with respect to the proper time has the 
eigenvalue
\begin{equation}
\ii \frac{\partial}{\partial \tau} \psi =
\frac{\ii}{\sqrt{T}} \, \frac{\partial}{\partial t} \psi 
= \frac{E}{\sqrt{T}} \, \psi = E_c \, \psi \,.
\end{equation}
All factors that go with $\sqrt{T}$ are unobservable since
they can be absorbed in going to local, Cartesian
coordinates. 

It is highly instructive (and non-obvious)
to convince oneself that the
leading kinetic terms in the Dirac--Schwarzschild
Hamiltonian~\eqref{HDSFW}
follow the $\sqrt{T}$ scaling. 
This observation, in particular, implies that the 
gravitational Breit term
\begin{equation}
- \frac{3 r_s}{8 m} \,
\left\{ \vec p^{\,2}, \frac{1}{r} \right\} \,,
\end{equation}
does not lead to an observable gravitational 
shift. At face value, one could otherwise assume
that it induces a numerically large, 
$(1/n^2)$-dependent shift 
on hydrogen energy levels
(where $n$ is the principle quantum number), because the 
operator $1/r$, where $r$ is the radial variable 
with respect to the gravitational center (e.g., the Earth), 
commutes, to an excellent approximation, with the 
momentum operator of the electron,
and in fact, the difference of the 
operator $(1/r) \, \vec p^{\,2}$ and the 
anti-commutator $(1/2) \, \{ r^{-1}, \vec p^{\,2} \}$ 
can be ignored altogether on the level 
of first-order perturbation theory.
This is because one has 
$\langle \psi^\plus | r^{-1} \, \vec p^{\,2} | \psi \rangle = 
\frac12 \, \langle \psi^\plus | \{ r^{-1}, \, \vec p^{\,2} \} | \psi \rangle$
for any reference state $\psi$.
(The Hermitian adjoint, as opposed to the Dirac 
adjoint $\overline \psi$, is denoted as $\psi^\plus$.)

The kinetic terms from Eq.~\eqref{HDSFW} 
read as follows,
\begin{align}
\label{Hkin_spec}
H_{\rm kin} = & \;
m - \frac{r_s}{2 \, r} \, m
+ \frac{\vec p^{\,2}}{2 m} 
- \frac{3 r_s}{8 m} \, \left\{ \vec p^{\,2}, \frac{1}{r} \right\}
\nonumber\\[0.1133ex]
\to & \;
m \left(1 - \frac{r_s}{2 \, r} \right) + 
\left( 1 - \frac{3}{2} \, \frac{r_s}{r} \right) \,
\frac{\vec p^{\,2}}{2 m} \,,
\end{align}
where we ignore the commutator and specialize the 
Hamiltonian to particles as opposed to 
anti-particles (i.e., we replace the 
Dirac $\beta$ matrix by the unit matrix).
For a central gravitational field, one has
\begin{equation}
T = 1 + 2 \, \Phi = 1 - \frac{r_s}{r} \,,
\qquad
H = 1 - 2 \, \Phi = 1 + \frac{r_s}{r} \,,
\end{equation}
where $r_s = 2 G M$.
To first order in $r_s$, we can thus reformulate the 
gravitational dependence as follows,
\begin{equation}
\label{Hkin}
H_{\rm kin} \sim 
\sqrt{T} \, m + \frac{\sqrt{T}}{H} \,
\frac{\vec p^{\,2}}{2 m} =
\sqrt{T} \, \left( m + 
\frac{\vec p_c^{\,2}}{2 m} \right) \,.
\end{equation}
Here, we have transformed the momentum operator 
to local Cartesian coordinates, as follows,
\begin{equation}
p_c^j 
= -\ii \frac{\partial}{\partial x_c^j} 
= -\ii \frac{1}{\sqrt{H}} \frac{\partial}{\partial x^j} 
= \frac{1}{\sqrt{H}} p^j \,.
\end{equation}
This implies that the gravitational Breit 
term does not contribute to an observable 
gravitational energy difference among atomic 
energy levels. 

The Schr\"{o}dinger Hamiltonian is completed
by adding the Coulomb term
\begin{equation}
\label{Hcoul}
H_{\rm coul} = - \frac{Z e^2}{4 \pi \epsilon \, \rho} 
= - \sqrt{\frac{T}{H}} \frac{Z e^2}{4 \pi \, \rho}
= - \sqrt{T} \, \frac{Z e^2}{4 \pi \, \rho_c} \,,
\end{equation}
where $\epsilon = \sqrt{H/T}$ is the gravitationally 
modified vacuum permittivity,
$\rho = | \vec \rho |$ is the distance from the 
atomic nucleus, and $\rho_c = \sqrt{H} \, \rho$.
It is instructive to compare the scaling 
outlined above to the relativistic formalism
used in Sec.~\ref{sec2B}.

Adding the kinetic term from Eq.~\eqref{Hkin}
and the Coulomb term given in Eq.~\eqref{Hcoul},
and subtracting the rest mass term,
which is irrelevant for atomic transitions,
one obtains the gravitationally
modified Schr\"{o}dinger Hamiltonian
\begin{equation}
\label{HST}
H_S = \sqrt{T} \, \left(
\frac{\vec p_c^{\,2}}{2 m}
- \frac{Z \alpha }{\rho_c} \right) \,,
\end{equation}
where $\alpha = e^2/(4 \pi)$ is the fine-structure
constant.

Interestingly, one could hypothesize about the 
physical consequences of the gravitational Breit 
term for high-precision atomic clocks~\cite{HiEtAl2013,HuEtAl2016},
which currently operate on a precision level of $10^{-18}$ 
or better, if the Breit term were to contribute
to an observable energy difference and the 
Penrose conjecture were to hold in the renormalized 
form~\eqref{penrose_ren}.
In this case, the renormalized gravitational
energy difference~\eqref{penrose_ren}
among the different atomic levels
involved in the atomic clock transition,
in view of $r_s/r \sim 10^{-9}$ for the Earth,
would result in gravitational collapse of the
atomic state on a relative frequency level of 
$10^{-9}$, which is the ratio of the 
gravitational Breit term to the 
leading nonrelativistic kinetic term in the atomic 
Hamiltonian [$\vec p^{\,2}/(2m)$].
This would prevent continuous 
interrogation of the atomic clock and thus make
the experiments~\cite{HiEtAl2013,HuEtAl2016} 
(and also the clock comparison, 
see Refs.~\cite{GoEtAl2014,HuEtAl2014}) infeasible,
because the hypothetical gravitational effect
would limit the clock precision to a 
level of $10^{-9}$. Indeed, the Yb$^+$ clock 
transitions described in Refs.~\cite{HiEtAl2013,HuEtAl2016}
involve atomic transitions with a change in the 
principal quantum number, and thus, the expectation value
of the  $\vec p^{\,2}$ operator becomes state dependent.
This hypothetical consideration is included here
in order to illustrate that care is required in the 
treatment of the gravitational terms;
one can easily be fooled into obtaining 
excessively large effects if one does not 
carry through the analysis correctly
(see also 
Refs.~\cite{MoFuSh2018ptep,MoFuSh2018remark1,MoFuSh2018remark2,Vi2018,Ni2018,Gu2018}).

\begin{table*}[ht!]
\begin{center}
\begin{minipage}{0.9\linewidth}
\begin{center}
\caption{\label{table1} Order-of-magnitude estimate
(rows 1, 2, 4, 5) and numerical values (remaining rows)
of the quantum limitations of the EEP, for the effects 
$\delta E^{(i)}$, $\delta E^{(ii)}$, $\delta E^{(iii)}$
$\delta E^{(iv)}$ as described in the text
[see Eqs.~\eqref{dEi},~\eqref{dEii}, and~\eqref{dEiii}].
The shifts are 
evaluated for astrophysical objects of interest.
The column labeled ``Earth due to Sun'' is 
included because, despite the large distance 
from the Earth to the Sun (about $146 \times 10^9 \, {\rm m}$),
the large solar mass of about $M_\odot = 1.989 \times 10^{30} \, {\rm kg}$ 
could be assumed to lead to large 
gravitational shifts. However, because of the 
suppression of the gravitational effects by
$R^{-n}$, with $n \geq 2$, the effects due to the Sun 
are numerically suppressed.}
\begin{tabular}{l@{\hspace*{0.3cm}}S[table-format=-1.2e-2]@{\hspace*{0.3cm}}%
S[table-format=-1.2e-2]@{\hspace*{0.3cm}}S[table-format=-1.2e-2]%
@{\hspace*{0.3cm}}S[table-format=-1.2e-2]}
\hline
\hline
\multicolumn{1}{c}{Effect} & 
\multicolumn{1}{c}{Earth} & 
\multicolumn{1}{c}{Earth} & 
\multicolumn{1}{c}{White} & 
\multicolumn{1}{c}{Neutron} \\
& & \multicolumn{1}{c}{Due to Sun} & 
\multicolumn{1}{c}{Dwarf} & 
\multicolumn{1}{c}{Star} \\
\hline
\stdrule  
$\delta E^{(i)}/E_h$ [Eq.~\eqref{dEi}, estimate]
& 8.99e-40 & 2.50e-47 & 4.19e-34 & 2.72e-26 \\
\hline
\stdrule
$\delta E^{(ii)}/E_h$ [Eq.~\eqref{dEii}, estimate] & 
1.17e-44 & 4.74e-51 & 2.55e-33 & 1.06e-22 \\
\stdrule
$\delta E^{(ii)}/E_h$ [Eq.~\eqref{dEii_calc}, hydrogen~$2S$] &
1.77e-37 & 7.13e-44 & 3.84e-26 & 1.59e-15 \\
\hline
\stdrule
$\delta E^{(iii)}/E_h$ [Eq.~\eqref{dEiii}, estimate] & 
4.79e-44 & 1.33e-51 & 2.23e-38 & 1.45e-30 \\
\hline
\stdrule
$\delta E^{(iv)}/E_h$ [Eq.~\eqref{Eiv_estim}, estimate] & 
1.99e-19 & 1.26e-22 & 9.28e-14 & 1.89e-8 \\
\stdrule
$\delta E^{(iv)}/E_h$ [Eq.~\eqref{Eiv_calc}, HF] &
3.17e-20 & 2.02e-23 & 1.48e-14 & 3.01e-9 \\
\stdrule
$\delta E^{(iv)}/E_h$ [Eq.~\eqref{Eiv_calc}, N$_2$] &
8.89e-19 & 5.65e-22 & 4.14e-13 & 8.43e-8 \\
\stdrule
$\delta E^{(iv)}/E_h$ [Eq.~\eqref{Eiv_calc}, Cl$_2$] &
-1.32e-18 & -8.41e-22 & -6.17e-13 & -1.25e-7 \\
\hline
\hline
\end{tabular}
\end{center}
\end{minipage}
\end{center}
\end{table*}

%
%
\subsection{Higher orders
and broken $\maybebm{\sqrt{T}}$ scaling}
\label{sec3B}

%
%
\subsubsection{Overview}
\label{sec3B1}

From this consideration, 
it becomes obvious that only gravitational
effects on atomic transitions which go beyond the
``common prefactor'' $\sqrt{T}$
[see Eq.~\eqref{common_prefactor}], could lead to 
observable consequences (competing effects
are discussed in Apps.~\ref{appB} and~\ref{appC}).
We therefore attempt, for a central gravitational field,
to analyze the leading effects which could 
contribute to quantum limitations of the EEP,
in view of a breaking of the $\sqrt{T}$ scaling.
There are three competing effects
to compare and to analyze,
{\em (i)} a first-order plain gravitational shift,
obtained by expanding the Newtonian gravitational
potential over the size of the atom,
{\em (ii)} a second-order gravitational shift,
again obtained on the basis of 
the Newtonian gravitational potential,
{\em (iii)} 
commutator-induced shifts due to higher-order operators
in the Dirac--Schwarzschild--Coulomb Hamiltonian.
A fourth effect, quite surprisingly,
exists for diatomic molecules.

%
%
\subsubsection{Gravitation and size of the atom}
\label{sec3B2}

We denote by $\vec R$ the coordinate of
the atomic nucleus with respect to the 
gravitational center, and by $\vec \rho$ the 
distance of the electron from the atomic nucleus.
Then, if $\vec \rho$ denotes the vector 
from the gravitational center to the atomic electron,
one has
\begin{align}
V =& \; - \frac{m \, r_s}{2 \, r}
= m \, \Phi = -\frac{G m M}{r} \,,
\nonumber\\[0.1133ex]
\frac{1}{r} = & \;
\frac{1}{| \vec R + \vec \rho | } = 
\frac{1}{R} - \frac{\vec R \cdot \vec \rho}{R^3} 
+ \frac{3 \, (\vec R \cdot \vec \rho)^2 - R^2 \, \rho^2}{2 \, R^5} 
\nonumber\\[0.1133ex]
& \; - \vec \rho^{\,2} \, 4 \pi \delta^{(3)}(\vec R) +
\calO(\rho^3) \,,
\end{align}
where the Dirac-$\delta$ term can be ignored 
if the atom is sufficiently 
displaced from the point $\vec R = \vec 0$,
which can be safely assumed to be 
the case for practically important applications.
One writes
\begin{subequations}
\begin{align}
V =& \; V^{[0]} + V^{[1]}  + V^{[2]} \,,
\\[0.1133ex]
V^{[0]} =& \; -\frac{G m M}{R} \,,
\\[0.1133ex]
V^{[1]} =& \; G m M \frac{\vec R \cdot \vec \rho}{R^3}
\propto \Phi^2 \,,
\\[0.1133ex]
V^{[2]} =& \; - \frac{G m M}{2}
\frac{3 \, (\hat R \cdot \vec \rho)^2 - \vec \rho^2}{R^3}  
\propto \Phi^3 \,.
\end{align}
\end{subequations}
The term $V^{[0]}$ is absorbed in the scaling factor $\sqrt{T}$
which multiplies the mass term in $H_{\rm kin}$,
as given in Eqs.~\eqref{Hkin_spec} and~\eqref{Hkin}.
The expectation value of the leading correction $V^{[1]}$
vanishes on any atomic energy eigenstate, due to parity.
However, as shown in the following,
nontrivial effects can be expected for diatomic molecules.
The effect scales as $R^{-2}$ and thus is 
proportional to $\Phi^2$, where $\Phi = -G M/R$ is the gravitational 
potential.

The first nonvanishing correction is due to the 
quadrupole term $V^{[2]}$, which scales with $\Phi^3$.
For an atom, $|\vec \rho| \sim a_0$ where $a_0$ is the Bohr radius.
The induced shift is of order 
\begin{equation}
\label{dEi}
\delta E^{(i)} = \langle V^{[2]} \rangle \sim
\frac{G \, m \, M \, a_0^2}{R^3} 
= 8.99 \times 10^{-40} \, E_h \,,
\end{equation}
where $E_h = \alpha^2 m c^2 \approx 27.2 \, {\rm eV}$ is the Hartree energy
and the shift has been evaluated for the Earth
($M \to M_\oplus$, $R \to R_\oplus$).
For other systems, see Table~\ref{table1}.
The effect is of first order in the 
gravitational potential and addresses 
point {\em (i)} listed above.

For completeness, we should point out that
we use the following parameters: the Earth mass
$M_\plus = 5.974 \times 10^{24} \, {\rm kg}$,
the Sun's mass $M_\odot = 1.989 \times 10^{30} \, {\rm kg}$,
a typical white dwarf mass of $M_{\rm wd} = 1.4 \, M_\odot$,
with a radius of $R_{\rm wd}$ being equal to the
radius of the Earth, $R_{\rm wd} = R_\oplus = 6.378 \times 10^6 \, {\rm m}$,
as well as a neutron star of mass $M_{\rm ns} = 2.8 \, M_\odot$,
and a radius of $R_{\rm ns} = 20 \, {\rm km}$.

The second-order perturbation due to $V^{[1]}$, on an atomic state, 
can be expressed as
\begin{align}
\label{dEii}
\delta E^{(ii)} =& \; \left< V^{[1]} \, \frac{1}{(E - H)'} \, V^{[1]} \right> 
\nonumber\\[0.1133ex]
\sim & \; 
\frac{G^2 \, m \, M^2 \,a_0^2}{\alpha^2 \, c^2 \, R^4}
\sim 1.17 \times 10^{-44} \, E_h \,,
\end{align}
where $[1/(E- H)']$ is the atomic reduced Green function,
and the numerical value is obtained for a point on the 
surface of the Earth.
We here assume that there are no quasi-degenerate levels 
which are displaced from the atomic reference state 
by an energy shift which is far less than a typical 
atomic energy level difference of 
$E - E_n \sim E_h \equiv \alpha^2 \, m \, c^2$,
where $E$ is the reference-state energy, and 
$E_n$ is the virtual-state energy. 
The Hartree energy is denoted as $E_h \approx 27.2 \, {\rm eV}$.
In the absence of such quasi-degenerate levels,
the order-of-magnitude estimate
$[1/(E- H)'] \sim 1/E_h$ is valid.
One may consult Table~\ref{table1} 
for numerical estimates of $\delta E^{(ii)}$ for other astrophysical 
systems. We have thus addressed point {\em (ii)} listed above.

A remark is in order. 
The estimate given above in Eq.~\eqref{dEii}
should be taken with a grain of 
salt, in part, because quasi-degenerate levels
can otherwise alter the predictions quite drastically. 
E.g., for the hydrogen $1S$--$2S$ 
transition~\cite{FiEtAl2004,PaEtAl2011}, 
the $2P_{1/2}$ levels are displaced from the 
$2S$ state only by the Lamb shift, while
the $2P_{3/2}$ levels are separated by the fine structure.
With the following data 
[see Eq.~(42) of~\cite{AdEtAl2017vdWi}]
for the $2S$--$2P_{1/2}$ Lamb shift energy 
interval $\calL$ and the 
$2P_{3/2}$--$2P_{1/2}$ fine-structure
interval $\calF$,
\begin{align}
\calL =& \; 1.61 \times 10^{-7} \, E_h \,,
\\[0.1133ex]
\calF =& \; 1.67 \times 10^{-6} \, E_h \,,
\end{align}
we have
[see Eq.~(17) of~\cite{AdEtAl2017vdWi}],
\begin{equation}
\left< z\, \frac{1}{(E - H)'} \, z \right>
= 3 \, a_0^2 \, \left( \frac{1}{{\cal L}} - 
\frac{2}{{\cal F}} \right)\,.
\end{equation}
The estimates in the second row of
Table~\ref{table1} should thus be
multiplied by a factor
\begin{equation}
\label{dEii_calc}
3 \left( \frac{1}{\calL} - \frac{2}{\calF} \right) =
1.504 \times 10^7 \,,
\end{equation}
to obtain numbers for the hydrogen $2S$ state.
The modified estimates, 
adjusted for the hydrogen $2S$ state,
are given in the third row of Table~\ref{table1}.

\begin{table*}[th!]
\begin{center}
\begin{minipage}{0.9\linewidth}
\begin{center}
\caption{\label{table2} Order-of-magnitude estimate
(rows 1, 2, 4, 5) and numerical values (remaining rows) for the 
$C_n(M)$ coefficients for those gravitational shifts 
of atomic transitions
which break the $\sqrt{T}$ scaling, 
as defined in Eq.~\eqref{defCM}.}
\begin{tabular}{l@{\hspace*{0.3cm}}S[table-format=-1.2e-2]@{\hspace*{0.3cm}}%
S[table-format=-1.2e-2]@{\hspace*{0.3cm}}S[table-format=-1.2e-2]%
@{\hspace*{0.3cm}}S[table-format=-1.2e-2]}
\hline
\hline
\multicolumn{1}{c}{Effect} & 
\multicolumn{1}{c}{Earth} & 
\multicolumn{1}{c}{Earth} & 
\multicolumn{1}{c}{White} & 
\multicolumn{1}{c}{Neutron} \\
& & \multicolumn{1}{c}{Due to Sun} & 
\multicolumn{1}{c}{Dwarf} & 
\multicolumn{1}{c}{Star} \\
\hline
\stdrule
$C_3(M)$ for $\delta E^{(i)}$ [$\hbar \omega_0 = E_h$]
& 8.99e-40 & 8.11e-51 & 4.14e-51 & 1.03e-51 \\
\hline
\stdrule
$C_2(M)$ for $\delta E^{(ii)}$ [$\hbar \omega_0 = E_h$]
& 1.17e-44 & 2.24e-53 & 1.17e-44 & 1.19e-39 \\
\stdrule
$C_2(M)$ for $\delta E^{(ii)}$ [hydrogen~$2S$, $\omega_0 = \omega_{1S2S}$] &
4.71e-37 & 8.99e-46 & 4.71e-37 & 4.79e-32 \\
\hline
\stdrule
$C_3(M)$ for $\delta E^{(iii)}$ [$\hbar \omega_0 = E_h$]
& 4.79e-44 & 4.32e-55 & 2.20e-55 & 5.51e-56 \\
\hline
\stdrule
$C_2(M)$ for $\delta E^{(iv)}$ [$\hbar \omega_0 = E_h$]
& 1.99e-19 & 5.98e-25 & 4.27e-25 & 2.13e-25 \\
\stdrule
$C_2(M)$ for $\delta E^{(iv)}$ [HF, $\omega_0 = \omega_{\rm ioni}$]
& 1.41e-19 & 4.24e-25 & 3.03e-25 & 1.51e-25 \\
\stdrule
$C_2(M)$ for $\delta E^{(iv)}$ [N$_2$, $\omega_0 = \omega_{\rm ioni}$]
& 1.55e-18 & 4.66e-24 & 3.33e-24 & 1.66e-24 \\
\stdrule
$C_2(M)$ for $\delta E^{(iv)}$ [Cl$_2$, $\omega_0 = \omega_{\rm ioni}$]
& -3.14e-18 & -9.42e-24 & -6.73e-24 & -3.37e-24 \\
\hline
\hline
\end{tabular}
\end{center}
\end{minipage}
\end{center}
\end{table*}

%
%
\subsubsection{Fokker precession term}
\label{sec3B3}

The Fokker precession term 
\begin{equation}
H_{\rm FP} =
\frac{3 r_s}{8 m} \, \frac{\vec\sigma \cdot \vec L}{R^3} 
\end{equation}
in the Dirac--Schwarzschild--Coulomb Hamiltonian~\eqref{HDSFW}
is proportional to $|\Phi|^3$,
where $\Phi = -G M /r$ is the gravitational potential.
This term is generated by the difference of the
exact Foldy--Wouthuysen Hamiltonian, given in Eq.~\eqref{HDSFW},
and the approximate form~\eqref{HDSCproblem},
due to commutators of the momentum operators and the
gravitational potential, while
we had obtained the approximate form~\eqref{HDSCproblem}
by {\em ignoring} the commutators.
It thus goes beyond the terms considered in Ref.~\cite{Wi1974prd},
which exhibit the universal scaling with
$\sqrt{T} = \sqrt{1 + 2 \Phi}$,
and leads to an energy shift of the order of
\begin{equation}
\label{dEiii}
\delta E^{(iii)} = \langle H_{\rm FP} \rangle \sim
\frac{\hbar^2 G M}{m R^3 c^2} =
4.79 \times 10^{-44} \, E_h \,.
\end{equation}
The numerical estimate is obtained for the Earth
($M = M_\oplus$ and $R = R_\oplus$).
Again, one may consult Table~\ref{table1}
for numerical estimates of $\delta E^{(iii)}$ for other astrophysical
systems. We have addressed point {\em (iii)} listed above.

%
%
\subsubsection{Atoms and limit of vanishing Bohr radius}
\label{sec3B4}

One might argue that the variation
of the gravitational potential around the 
atomic center does not constitute a quantum limitation
of the EEP, because it is simply given
as the expectation value of 
a gravitational effect, evaluated on the 
atomic wave function.
However, it leads to an observable frequency shift
and to a deviation from the universal $\sqrt{T}$ 
scaling of the atomic transition frequencies.
The effect would vanish
if the electron could be perfectly localized, which 
however is incompatible with fundamental 
postulates of quantum mechanics. In particular,
perfect localization of the 
electron's wave packet 
would be incompatible with Heisenberg's uncertainty principle.
It is instructive to observe that the 
energy shifts $\delta E^{(i)}$ and $\delta E^{(ii)}$ 
vanish in the limit $a_0 \to 0$, which 
would correspond to the classical limit of a 
perfectly localizable electron.
The energy shifts $\delta E^{(iii)}$,
by contrast, is nonvanishing even in the 
limit $a_0$ and constitutes a genuine 
quantum correction to the EEP, due to 
the Fokker precession acting on the bound 
atomic electron. 

%
%
\subsubsection{Diatomic molecules}
\label{sec3B5}

For a diatomic molecule, the situation is essentially more
interesting because the expectation value
\begin{equation}
\label{Eiv}
\delta E^{(iv)} =
\langle V^{[1]} \rangle = 
\left< G m M \frac{\vec R \cdot \vec \rho}{R^3} \right>
\end{equation}
can be nonvanishing.
It is known that diatomic molecules typically have
nonvanishing electric dipole moments~\cite{HuHe1979}.
Indeed, it is known~\cite{TaSi1977} that, e.g.,
hydrogen fluoride (HF) has a dipole moment of
$1.82 \, {\rm D}$, where $\rm D$ denotes the Debye,
which is a canonical unit of an atomic dipole moment,
equal to $0.20819434 \, |e| \, \mbox{\AA}$,
where $|e|$ is the elementary charge.
A calculation using GAUSSIAN 2.0~\cite{Kl2018Priv}
reveals that the hydrogen fluoride ion (HF$^+$) 
has a dipole moment of $2.36 \, {\rm D}$
(in units of Debye), which is 
measured with respect to the center-of-mass
of the hydrogen fluoride ion
(by convention). However, the {\em electric} dipole moment is of 
no significance when it comes to the 
evaluation of gravitational corrections.

Namely, for the evaluation of gravitational corrections,
one should consider the fact that the 
mass of the atom is concentrated in the 
atomic nuclei. The two nuclei in a diatomic 
molecule are separated by the bond length.
If the energetically highest molecular orbital
is bonding, then the bond length will increase upon
excitation into energetically higher states,
with the maximum change reached for excitations
close to the ionization threshold.
An example is HF, which has a bond length of 
\begin{equation}
\label{ell_start}
\ell_{\rm HF} = 0.917\, \mbox{\AA},
\end{equation}
to be contrasted with HF$^+$, which has a 
bond length of 
\begin{equation}
\ell_{\rm HF^+} = 1.001\,\mbox{\AA} \,,
\end{equation}
according to Refs.~\cite{TaSi1977,HuHe1979}.

By contrast, if the energetically highest molecular orbital
is anti-bonding, then the bond length will decrease upon
excitation into energetically higher states,
An example is CL$_2$, whose bond length decreases
from 
\begin{equation}
\ell_{\rm CL_2} = 1.99\,\mbox{\AA}  \qquad \to \qquad
\ell_{\rm CL_2^+} = 1.89\,\mbox{\AA} 
\end{equation}
upon ionization into CL$_2^+$
(see Ref.~\cite{HuHe1979}).
For N$_2$, the bond length changes 
according to 
\begin{equation}
\label{ell_end}
\ell_{\rm N_2} = 1.12\,\mbox{\AA}  \qquad \to \qquad
\ell_{\rm N_2^+} = 1.29\,\mbox{\AA} \,.
\end{equation}

We can thus conclude that, in a diatomic molecule,
if we hold the position of one of the nuclei 
(mass $m_1$) fixed to the origin, then
there will be an energy correction of the form
\begin{align}
\label{Eiv_estim}
\delta E^{(iv)} =& \;
\langle V^{[1]} \rangle =
\left< G M \frac{m_2 \vec R \cdot \vec L}{R^3} \right> 
\nonumber\\[0.1133ex]
\sim & \;
\frac{G \, m_p \, M \, a_0}{R^2} \,.
\end{align}
Here, $\vec L$ is the bond length vector, 
$m_2$ is the mass of the respective other nucleus,
while $m_p$ is the proton mass.
In formulating the order-of-magnitude estimate,
we use the proton mass (mass of the nucleus of the hydrogen
atom) as a measure for $m_2$; of course, this 
assumption has to be adjusted according to 
the molecule under consideration.

The ionization energy of a diatomic molecule
in a gravitational field thus changes according to 
\begin{equation}
\label{Eiv_calc}
\delta E^{(iv)} = \frac{G m_2 M \, \Delta \ell}{R^2} 
\end{equation}
upon ionization, if the axis of the diatomic molecule
is aligned along the $\vec R$ vector.
Here, $\Delta \ell$ is the change in the bond 
length upon ionization.
This is because directly under the ionization 
threshold, the bond length will 
asymptotically approach that of the ion.

According to the above considerations, 
given in Eqs.~\eqref{ell_start}---\eqref{ell_end}, one has
\begin{subequations} 
\begin{align} 
\Delta \ell_{\rm HF} =& \; 0.084 \, \mbox{\AA}\,,
\\[0.1133ex]
\Delta \ell_{{\rm N}_2} =& \; 0.17 \, \mbox{\AA}\,,
\\[0.1133ex]
\Delta \ell_{{\rm Cl}_2} =& \; -0.10 \, \mbox{\AA}\,.
\end{align}
\end{subequations} 
Numerical estimates of the gravitational
effects can be quite large for typical diatomic 
For absolute clarity, we should
point out that, for a successful
measurement of the gravitation frequency shift, 
the diatomic molecules need to 
be aligned with reference to the gravitational field;
of course, the effect vanishes when averaged over an
ensemble of unaligned molecules.

%
%
\section{Measurement of the Higher--Order Shifts}
\label{sec4}

According to Table~\ref{table1},
the dominant effects for either the hydrogen 
$1S$--$2S$ transition or molecular 
transitions are given by the shifts 
$\delta E^{(ii)}$ and $\delta E^{(iv)}$.
It is instructive to study their
dependence on the gravitational potential,
and the measurability of the effects.

As evident from Eq.~\eqref{dEii}, the 
shift $\delta E^{(ii)}$ can be written as
\begin{equation}
\label{before}
\delta E^{(ii)} = 
\left( \frac{\Phi}{\Phi_0} \right)^4 \, 
C_4(M) \, (\hbar \omega_{1S2S}) \,,
\end{equation}
where $\omega_{1S2S}$ is the unperturbed $1S$--$2S$ 
frequency, and $C_4(M)$ is a coefficient whose value 
depends on the mass of the gravitational center.
The gravitational potential is $\Phi = -G M/R$,
and we have normalized the potential with respect to 
$\Phi_0 = G M_\oplus/R_\oplus$,
where $M_\oplus$ is the Earth mass,
and $R_\oplus$ is the Earth's radius.
Also, $\delta E^{(iv)}$ given in Eq.~\eqref{Eiv_estim}
can be written as
\begin{equation}
\delta E^{(iv)} = \left( \frac{\Phi}{\Phi_0} \right)^2 \, 
C_2(M) \, (\hbar \omega_{\rm ioni}) \,,
\end{equation}
where $\omega_{\rm ioni}$ is the 
angular unperturbed ionization frequency,
and $C_2(M)$ [which may be different from the coefficient
used in Eq.~\eqref{before}]
is a mass-dependent coefficient.

Let us assume that a general higher-order 
gravitational frequency shift,
which limit the validity of the universal $\sqrt{T} $ scaling,
can be written in the functional form
\begin{equation}
\label{defCM}
\delta E = 
\left( \frac{|\Phi|}{\Phi_0} \right)^n \, 
C_n(M) \, \hbar \omega_0 \,,
\end{equation}
where, for the cases studied above,
one would have either $n = 2,3,4$. 
The coefficient $C_n(M)$ depends on the 
mass of the gravitational center, 
while $\omega_0$ is the unperturbed frequency.

We can thus write a gravitationally corrected
transition energy $E$ as
\begin{align}
E =& \; \sqrt{T} \, \omega_0 + \delta E
\nonumber\\[0.1133ex]
=& \; \left( \sqrt{1 + 2 \Phi} +
| \Phi |^n \, C(M) \right) \, \hbar \omega_0 \,.
\end{align}
In units with $\hbar = 1$, we have 
\begin{equation}
E = \frac{\dd \theta}{\dd t} \,,
\qquad
\omega_0 = \frac{\dd \theta}{\dd \tau} \,,
\end{equation}
where $t$ is the global coordinate time,
and $\tau$ is the proper time measured by the local
observer, while $\theta$ is the rotation angle
of the oscillation. Then,
\begin{equation}
\frac{\dd \tau}{\dd t} = \sqrt{1 + 2 \Phi} +
| \Phi |^n \, C_n(M) \,.
\end{equation}
Comparing two atomic clocks at different 
altitudes (points labeled $1$ and $2$ in the 
gravitational field), which is the essence of relativistic 
geodesy~\cite{MaMu2013}, one arrives at the result
\begin{equation}
\label{corr_term}
\frac{\dd \tau_1}{\dd \tau_2} = 
\frac{\sqrt{1 + 2 \Phi_1} + | \Phi_1 |^n \, C_n(M)}%
{\sqrt{1 + 2 \Phi_2} + | \Phi_2 |^n \, C_n(M)} \,.
\end{equation}
We reemphasize that the numerical value of the 
coefficient $C_n(M)$, as well as the value of $n$, 
are not universal, but depend on the 
atomic system and the transition under study
(see Table~\ref{table2}).
The prediction thus is that, 
if one expands this result in $\Phi_1$ and $\Phi_2$ to order $n$, then
the coefficients of order less than $n$ will 
agree with the expansion of the leading term 
$\sqrt{1 + 2 \Phi_1}/\sqrt{1 + 2 \Phi_2}$, 
while at order $n$, there will be an additional correction
\begin{align}
\label{expansion}
\frac{\dd \tau_1}{\dd \tau_2} \sim & \;
\left(
\sum_{k=0}^n \left( \begin{array}{c}
\sfrac12 \\ k \end{array} \right) (2 \Phi_1)^k
\right)
\left(
\sum_{k=0}^n \left( \begin{array}{c}
\sfrac12 \\ k \end{array} \right) (2 \Phi_2)^k
\right)^{-1}
\nonumber\\[0.1133ex]
& \; + C_n(M) \, \left( | \Phi_1 |^n \, - | \Phi_2 |^n \right) \,,
\end{align}
which describes the deviation from Einstein's
equivalence principle.
Here, 
\begin{equation}
\left( \begin{array}{c}
n \\ k \end{array} \right) = 
\frac{\Gamma(n+1)}{\Gamma(k+1) \, \Gamma(n - k + 1)}
\end{equation}
is the binomial coefficient.
Terms of order $n+1$ and higher in the 
gravitational potentials have
been neglected in writing Eq.~\eqref{expansion}.
Numerical results for the coefficient $C_n(M)$
are given in Table~\ref{table2}.
We use the ionization energies 
$6.12\,{\rm eV}$ for HF,
$15.58 \, {\rm eV}$ for N$_2$,
and $11.48 \, {\rm eV}$ for Cl$_2$,
as well as the known $1S$--$2S$  frequency 
for hydrogen (see Ref.~\cite{PaEtAl2011}).

%
%
\section{Conclusions}
\label{sec5}

Let us summarize the main results of the 
current, lengthy, paper. 
We shall proceed section by section.

In Sec.~\ref{sec2},
we derive generally applicable Hamiltonians 
for the combined gravitational-electromagnetic interaction
in a central gravitational field,
which add relativistic corrections to the 
leading-order (nonrelativistic) result 
[see Eqs.~\eqref{HEM} and~\eqref{HEMplus}].
Furthermore, we show that the interplay of the 
gravitationally modified Dirac equation,
and the gravitationally modified vacuum permittivity 
and permeability, leads to a value of the fine-structure 
constant independent of gravity [see Eq.~\eqref{alphaconst}].
As a result, we confirm (see Ref.~\cite{Wi1974prd}) that 
atomic transition energies are (to an excellent approximation)
compatible with the equivalence principle
[see Eq.~\eqref{yes}].
We also derive a universal gravitational scaling 
for the electron's $g$ factor, including the 
bound-state corrections, and the anomalous magnetic moment
term [see Eq.~\eqref{gJ1S}].
Only a careful consideration of the transformation of the 
magnetic-field components from global coordinates to a local
Lorentz frame, restores the validity of the EEP 
[see Eq.~\eqref{ottifant}].

In Sec.~\ref{sec3}, we first discuss 
gravitational energy shifts which scale with the 
universal prefactor $\sqrt{T} = \sqrt{1 + 2 \, \Phi}$.
Our discussion culminates in Eq.~\eqref{HST},
where we derive the gravitationally 
corrected Schr\"{o}dinger--Coulomb Hamiltonian,
to complement Eq.~\eqref{yes}.
Furthermore, we treat
four effects which go beyond the 
universal prefactor $\sqrt{T}$,
and which, therefore, in the 
language of Ref.~\cite{Wi1974prd},
limit the compliance of transition frequencies
with the Einstein equivalence principle.
These effects are mainly caused by the 
non-deterministic nature of quantum mechanics,
which prevents us from perfectly localizing an 
electron at a given point in time,
as described by Heisenberg's uncertainty principle.
Specifically, we have an energy correction $\delta E^{(i)}$,
due to a quadrupole term in the gravitational field,
given in Eq.~\eqref{dEi}, which leads to 
a nontrivial effect due to 
the nonvanishing extent of the quantum mechanical
wave function. A second correction
$\delta E^{(ii)}$ is due to a second-order effect
involving the dipole expansion about the 
gravitational center of the atom,
and $\delta E^{(iii)}$ is described by 
the Fokker precession term.
One notices that the energy shift $\delta E^{(iii)}$ does not vanish
in the limit $a_0 \to 0$.
The effect thus does not require the 
gravitational field to change significantly over the 
dimension of the atom, at variance with a remark
issued in the text following Eq.~(12.13) of Ref.~\cite{Pa1980prd}.
Then, for diatomic molecules, quite remarkably,
the dipole term $\delta E^{(iv)}$ due to first-order 
perturbation theory
involves the dipole expansion about the 
gravitational center of the atom;
its expectation value does not vanish and leads to a direction-dependent
energy shift. Numerical values 
for the energy shifts $\delta E^{(i)}$,
$\delta E^{(ii)}$, $\delta E^{(iii)}$, and $\delta E^{(iv)}$,
are given in Table~\ref{table1}.

In Sec.~\ref{sec4}, we discuss the 
measurability of the gravitational shifts
in atomic-clock comparisons.
One first observes that the energy shifts $\delta E^{(i)}$,
$\delta E^{(ii)}$, $\delta E^{(iii)}$, and $\delta E^{(iv)}$,
which limit the validity of the $\sqrt{T}$ scaling,
have a functional $| \Phi |^n$ dependence, where $\Phi$ is the 
gravitational potential.
They thus lead to a correction term 
in the atomic-clock comparison, as given in
in Eq.~\eqref{corr_term}, 
which could in principle be measured in an 
accurate comparison of atomic clocks 
running at places with different gravitational potentials.
Equation~\eqref{corr_term} is one of the main
results of the current paper.
Data for the $C_n(M)$ coefficients,
which enter Eq.~\eqref{defCM},
are given in Table~\ref{table2}.

One should remember that the conclusions of 
Ref.~\cite{Wi1974prd} crucially depend on the approximation that
commutator terms between the gravitational couplings and the kinetic operators
in the Hamiltonian can be neglected. Only under this assumption can 
the fundamental $\sqrt{T}$ scaling of the atomic energy levels be
derived. Here,
we go beyond this approximation and quantify those effects which do not
follow the universal $\sqrt{T}$ scaling.
We reemphasize that the Fokker precession term 
does not vanish in the limit of a pointlike atom (vanishing Bohr radius),
and leads to a manifest deviation 
of the gravitational modification of 
atomic transition frequencies from the fundamental $\sqrt{T}$ scaling,
which is otherwise crucial in establishing the compatibility 
of high-precision spectroscopy experiments
with the equivalence principle~\cite{Wi1974prd}.

The tiny gravitational corrections beyond the 
$\sqrt{T}$ scaling should be compared to 
effects due to space-time noncommutativity~\cite{SeWi1999,ChSJTu2001,DeEtAl2017}
(see App.~\ref{appB}),
and a conceivable limitation of the 
achievable accuracy due to a gravitationally 
induced collapse of the wave function
(Penrose conjecture, see Refs.~\cite{Pe1996penrose,Pe1998penrose,Pe2014penrose},
see App.~\ref{appC}).
The conclusion is that under reasonable assumptions,
they do not preclude the measurability of the 
quantum corrections outlined in Eqs.~\eqref{corr_term} and~\eqref{defCM},
as explained in detail in App.~\ref{appB3} and App.~\ref{appC3}.
In view of seemingly unstoppable progress in high-precision
spectroscopy~\cite{PrEtAl2013}, the effects could be 
of phenomenological relevance sooner than otherwise expected.

%
%
\centerline{\bf Acknowledgements}

The author acknowledges helpful conversations with 
Professor Sir Roger~Penrose, as well as
Professors Clifford M.~Will, Ulrich Bonse and Gregory S.~Adkins.
The formalism outlined in 
Ref.~\cite{Wi1974prd} has been an utmost important 
inspiration for our studies. This research was 
supported by the National Science Foundation (grant PHY--1710856)
and by the Missouri Research Board.

\appendix

%
%
\section{Theoretical Background}

%
%
\subsection{Dirac Hamiltonian and Hermiticity}
\label{appA1}

We set, with Ref.~\cite{JeNo2013pra},
\begin{equation}
T = w^2, \qquad H = v^2 \,,
\qquad
\dd s^2 = w^2 \dd t^2 - v^2 \dd \vec r^{\,2} \,.
\end{equation}
It is known from the literature 
(see, e.g., Refs.~\cite{Pa1980prd,Ob2001,JeNo2013pra})
that the Hamiltonian obtained from the 
variation of the fully relativistic curved-space action of a 
Dirac particle is not Hermitian [see Eq.~(11) of Ref.~\cite{JeNo2013pra}].
It reads as follows,
\begin{equation}
\label{calHDS}
\calH = \frac{w}{v} \vec\alpha \cdot \vec p - 
\frac{\ii}{2 v} \, \vec\alpha \cdot \vec \nabla w -
\frac{\ii w}{v^2} \vec\alpha \cdot \vec\nabla v +
\beta m w \,.
\end{equation}
Let us carry out the transformation which leads to 
a Hermitian operator, in great detail. One sets
\begin{equation}
H_0 = \frac{w}{v} \vec\alpha \cdot \vec p  \,.
\end{equation}
Then,
\begin{widetext}
\begin{align}
\label{X}
X =& \; v^{3/2} H_0 v^{-3/2} 
- \frac12 \, \left\{ \vec\alpha \cdot \vec p, \frac{w}{v} \right\}
\nonumber\\[0.1133ex]
=& \; 
v^{3/2} \frac{w}{v} v^{-3/2} \vec\alpha \cdot \vec p 
+ v^{3/2} \frac{w}{v} 
\left[ \vec\alpha \cdot \vec p, v^{-3/2} \right]
- \frac{w}{v} \vec\alpha \cdot \vec p
- \frac12 \, \left[ \vec\alpha \cdot \vec p, \frac{w}{v} \right]
\nonumber\\[0.1133ex]
=& \; 
v^{3/2} \frac{w}{v} \left[ -\ii \vec\alpha \cdot 
\left( -\frac32\, v^{-5/2} \right) \vec\nabla v \right]
- \frac12 \, \left( -\ii \frac{1}{v} \vec\alpha \cdot \vec\nabla w
+ \left[ -\ii \left( -\frac{w}{v^2} \right) \,
\vec\alpha \cdot \vec\nabla v \right] \right)
\nonumber\\[0.1133ex]
=& \;
\frac{3 \ii w}{2 v^2} \vec\alpha \cdot \vec\nabla v 
+ \frac{\ii}{2 v} \vec\alpha \cdot \vec\nabla w
- \frac{\ii w}{2 v^2} \vec\alpha \cdot \vec\nabla v 
= \frac{\ii w}{v^2} \vec\alpha \cdot \vec\nabla v
+ \frac{\ii}{2 v} \vec\alpha \cdot \vec\nabla w \,.
\end{align}
\end{widetext}
The last expression is easily
recognized as the negative sum of the second and
third terms in Eq.~\eqref{calHDS}, and, hence, one obtains
the relativistic and Hermitian
Dirac--Schwarzschild Hamiltonian~\cite{JeNo2013pra},
\begin{equation}
H_{\rm DS} = v^{3/2} \; \calH \; v^{-3/2} =
\frac12 \, \left\{ \vec\alpha \cdot \vec p, \frac{w}{v} \right\} + 
\beta m w \,.
\end{equation}
The original Hamiltonian $\calH$ can thus be written 
as follows,
\begin{equation}
\calH = v^{-3/2} \; H_{\rm DS} \; v^{3/2} \,.
\end{equation}
This relation, in particular, implies that 
$\calH$ and $H_{\rm DS}$ have the same eigenvalues.
In order to see this, consider an eigenfunction
$\Psi$ of $H_{\rm DS}$, with $H_{\rm DS} \; \Psi = E \, \Psi$.
The corresponding eigenstate of 
$\calH$ is $\Phi = v^{-3/2} \, \Psi$, with the 
same energy eigenvalue $E$. Hence, 
$\calH$ and $ H_{\rm DS}$ can be used
interchangeably in eigenvalue perturbation 
theory, a fact which has implicitly been 
used in Eq.~(13) of Ref.~\cite{Ob2001} and 
elucidated in greater detail in 
Refs.~\cite{JeNo2013pra}.
The equivalence of the eigenvalues also is
used throughout the current paper.

The Hermitian adjoint of $\calH$ is
\begin{equation}
\label{calHplus}
\calH^\plus 
= v^{3/2} \; H_{\rm DS} \; v^{-3/2} 
= v^3 \; \calH \; v^{-3} \,,
\end{equation}
and thus, not equal to $\calH$ itself.
Rather, we have the relation $\calH^\plus \, v^3 = v^3 \, \calH$.
The relation~\eqref{calHplus} is reminiscent 
of pseudo--Hermiticity, a property which 
has been discussed by Pauli~\cite{Pa1943}
and recently used in the analysis of 
a number of quantum systems~\cite{BeBo1998,Be2005,Mo2002i,Mo2002ii,Mo2002iii,%
JeWu2012epjc,JeWu2012jpa,NoJe2015tach}.

Let us now consider the a general matrix element
$\langle \psi | v^3 | \phi \rangle$ between 
general states $\psi$ and $\phi$ which fulfill the
time-dependent Schr\"{o}dinger equation 
$\calH \psi = \ii \partial_t \psi$ and
$\calH \phi = \ii \partial_t \phi$, for 
$\calH$ (not $H_{\rm DS}$!), 
\begin{align}
\ii \partial_t \langle \psi | v^3 | \phi \rangle =& \;
\langle \psi | v^3 | \ii \partial_t \phi \rangle -
\langle \ii \partial_t \psi | v^3 | \phi \rangle 
\nonumber\\[0.1133ex]
=& \; \langle \psi | v^3 | \calH \, \phi \rangle -
\langle \calH \psi | v^3 | \phi \rangle 
\nonumber\\[0.1133ex]
=& \; \langle \psi | v^3 \, \calH | \phi \rangle -
\langle \psi | \calH^\plus \, v^3 | \phi \rangle = 0 \,,
\end{align}
where we have used Eq.~\eqref{calHplus}.
Hence, we have shown that the generalized 
scalar product $\langle \psi | v^3 | \phi \rangle$ is 
conserved under the time evolution 
induced by $\calH$. This makes perfect sense 
in a metric $\dd s^2 = w^2 \dd t^2 - v^2 \dd \vec r^2$,
where $| \dd s | = v \, | \dd \vec r |$ for $\dd t = 0$.

\begin{widetext}

%
%
\subsection{Alternative form of the Dirac Hamiltonian}
\label{appA2}

In order to make a comparison with the literature,
let us try to compare the Dirac--Schwarzschild Hamiltonian
to the result given in Eq.~(42) of Ref.~\cite{IvWe2015},
which is formulated using a general potential
$U_+ = U_- = U = -G \, M/r$ [see Eq.~(2) of Ref.~\cite{IvWe2015}],
where we set the contribution of the chameleon field
discussed in Ref.~\cite{IvWe2015}
to zero. First of all, an important identity is 
\begin{equation}
\{ A, \{ A, B \} \} =
2 \{ A^2, B \} - [ A, [A, B]] \,.
\end{equation}
Here, we use this identity for $B = U$ and $A = \vec\nabla$,
and obtain for general $U$, 
\begin{align}
\{ \vec\nabla, \{ \vec\nabla, U \} \} =& \;
2 \{ \vec\nabla^2, U \} - \vec\nabla^2(U)
= 2 U \vec\nabla^2 +
2 \vec\nabla^2 U - \vec\nabla^2(U)
\nonumber\\[0.1133ex]
=& \; 2 U \vec\nabla^2 +
2 \vec\nabla [\vec \nabla, U] +
2 \vec\nabla U \cdot \vec \nabla
- \vec\nabla^2(U)
\nonumber\\[0.1133ex]
=& \; 2 U \vec\nabla^2 +
2 \vec\nabla [\vec \nabla, U] +
2 [\vec\nabla, U] \cdot \vec \nabla +
2 U \, \vec\nabla^2 - \vec\nabla^2(U)
\nonumber\\[0.1133ex]
=& \; 2 U \vec\nabla^2 +
2 [\vec\nabla, [\vec \nabla, U]] +
2 [\vec \nabla, U] \cdot \vec \nabla +
2 [\vec\nabla, U] \cdot \vec \nabla +
2 U \, \vec\nabla^2 - \vec\nabla^2(U)
\nonumber\\[0.1133ex]
=& \; 4 U \vec\nabla^2 +
4 \vec \nabla(U) \cdot \vec \nabla + \vec\nabla^2(U) \,.
\end{align} 
The result given in Eq.~(42) of Ref.~\cite{IvWe2015} 
can then be rewritten as follows,
under the identification $\beta = \gamma^0$,
\begin{align}
{\rm H}_3 =& \;
\beta \, \left( m - \frac{1}{2 m} \, \vec\nabla^2 + m U \right)
+ \frac{\beta}{2 m} \, \left( 
- 3 U \, \vec\nabla^2 
- 3 \vec\nabla(U) \cdot \vec\nabla 
- \frac34 \vec\nabla^2(U) \cdot \vec\nabla
\right)
- \frac{\beta}{4m} 
\ii \vec\Sigma \cdot [ 3 \vec\nabla U \times \vec\nabla ]
\nonumber\\[0.1133ex]
=& \;
\beta \, \left( m - \frac{1}{2 m} \, \vec\nabla^2
- \frac{G m M}{r} \right)
- \frac{3 \beta}{8 m} \, \{ \vec\nabla, \{ \vec\nabla, U \} \}
+ \frac{3 \beta}{4m}
\vec\Sigma \cdot \left[ \frac{ G M \vec r}{r^3} \times
(-\ii \vec\nabla) \right]
\nonumber\\[0.1133ex]
=& \;
\beta \, \left( m + \frac{\vec p^{\,2}}{2 m}
- \frac{G m M}{r} \right)
+ \frac{3 \beta}{8 m} \, 
\left\{ \left\{ -\frac{G M}{r}, \vec p \right\}, \vec p \right\}
+ \frac{3 \beta G M}{4 m r^3}
\vec\Sigma \cdot \vec r \times \vec p
\nonumber\\[0.1133ex]
=& \;
\beta \, \left( m + \frac{\vec p^{\,2}}{2 m}
- \frac{m r_s}{2 r} \right)
- \frac{3 \beta}{16 m} \, \left\{ \left\{ \frac{r_s}{r}, \vec p \right\}, \vec p \right\}
+ \frac{3 \beta r_s}{8 m r^3}
\vec\Sigma \cdot \vec r \times \vec p
\nonumber\\[0.1133ex]
=& \;
\beta \, \left( m + \frac{\vec p^{\,2}}{2 m}
- \frac{m r_s}{2 r} \right)
- \frac{3 \beta}{8 m} \, \left\{ \frac{r_s}{r}, \vec p^{\,2} \right\}
+ \frac{3 \beta \pi r_s}{4 m} \, \delta^{(3)}(\vec r) 
+ \beta \frac{3 r_s \, \vec\Sigma \cdot \vec L}{8 m r^3} \,.
\end{align}
Thus, the result given in Ref.~\cite{IvWe2015},
upon setting the chameleon field to zero,
is seen to be equivalent to the Dirac--Schwarzschild 
Hamiltonian~\cite{JeNo2013pra}.
However, it is also clear that the result given in Ref.~\cite{IvWe2015}
concerns a (chameleon-field inspired 
generalization of) the Dirac--Schwarzschild Hamiltonian,
but does not consider the Coulomb-field terms which 
must be added to obtain the Dirac--Schwarzschild--Coulomb Hamiltonian
discussed in Sec.~\ref{sec2}.
We also mention the necessity of adding the 
relativistic $\vec p^{\,4}$ correction, depending on the 
approximations used in a particular treatment of the 
problem.

\end{widetext}

%
%
\section{Penrose Conjecture}
\label{appB}

%
%
\subsection{Theoretical foundations}
\label{appB1}

The Copenhagen interpretation of quantum 
mechanics and the pertinent collapse of the 
wave function still give rise to interesting 
questions about the foundations of physical theory,
as discussed by Penrose~\cite{Pe1996penrose,Pe1998penrose,Pe2014penrose}.
The Penrose conjecture 
implies that gravity yanks objects back into a single location, without
any need to invoke observers or parallel universes.
The gravitationally induced effects 
envisaged by the Penrose conjecture 
should be compared to the quantum effects 
discussed in Sec.~\ref{sec3} of this article.

In Fig.~5 of Ref.~\cite{Pe2014penrose}, Penrose 
conjectures that collapse of the wave 
function to one of two possible 
states is induced on a time scale
\begin{equation}
\label{tC}
t_C \sim \frac{\hbar}{E_G} \,,
\end{equation}
where $E_G$ is the gravitational 
self-energy of the difference between the 
two mass distributions, which, notably,
is not equal to the difference of their gravitational 
self-energies.

According to the an (unnumbered) 
equation on p.~595 of Ref.~\cite{Pe1996penrose},
the relevant expression is 
\begin{equation}
\label{EG1}
E_G = - G \, \int \dd^3 x \int \dd^3 y \,
\frac{\left[ \rho(\vec x) - \rho'(\vec x) \right] \,
\left[ \rho(\vec y) - \rho'(\vec y) \right]}%
{| \vec x - \vec y |} \,,
\end{equation}
where $\rho(\vec r)$ and $\rho'(\vec r)$ are the 
two mass distributions.

Let us confront this expression with the 
well-known Colella--Overhauser--Werner 
experiment~\cite{OvCo1974,CoOvWe1975,BoWr1983,BoWr1984},
where, according to the experimental description in 
Ref.~\cite{CoOvWe1975}, neutrons are separated across an interferometer
with a side length of about 2.5\,cm, and an opening angle of 
$22.1^\circ$.
During the experiment, the neutrons are ``gravitationally 
bound to the Earth''.
It is important to analyze the predictions of
the Penrose conjecture for this experiment,
because a conceivable
collapse of the wave would otherwise preclude
the observation of interference fringes
in the experiment~\cite{CoOvWe1975}.

Let us denote the mass distribution of the Earth
by  $\rho_\oplus(\vec r)$.
We associate the neutron wave function with a mass distribution
$m_n \, f(\vec r - \vec r_n) $, where $m_n$ is the 
neutron mass and $f$ is a properly normalized sampling 
function, centered about the origin, and $\vec r_n$ 
is a point on the ``lower'' arm of the quantum interferometer.
The other state, which is part of the superposition
and is centered around the ``higher'' arm of the 
interferometer, has a mass distribution given by the sum of
the mass distribution of the Earth, $ \rho_\oplus(\vec r) $, 
and a neutron wave function with a mass distribution
$m_n \, f(\vec r - \vec r_n - \vec h) $, where $m_n$ is the 
neutron mass, and $\vec h$ is the vector that describes
the height difference of the ``elevation'' of the neutrons
in the gravitational field of the Earth.
In this case, 
\begin{subequations}
\begin{align}
\rho(\vec r) =& \; \rho_\oplus(\vec r) + m_n \, f(\vec r - \vec r_n) 
\nonumber\\[0.1133ex]
=& \; \rho_\oplus(\vec r) + {\widetilde \rho}(\vec r) \,,
\\[0.1133ex]
\rho'(\vec r) =& \;
\rho_\oplus(\vec r) + m_n \, f(\vec r - \vec r_n - \vec h) 
\nonumber\\[0.1133ex]
= & \; \rho_\oplus(\vec r) + {\widetilde \rho}'(\vec r) \,,
\\[0.1133ex]
\rho(\vec r) - \rho'(\vec r) =& \;
m_n \, [ f(\vec r - \vec r_n) - f(\vec r - \vec r_n - \vec h) ] \,,
\end{align}
\end{subequations}
where $\rho_\oplus(\vec r)$ is the mass density 
of the Earth, and $m_n$ is the neutron mass, while
$f$ is normalized according to
$\int \dd^3 r \, f(\vec r) = 1$.
One might assume that $f \sim | \psi |^2$, where $\psi$ is the 
quantum mechanical wave function.

In this case, the expression~\eqref{EG1} can be written 
as the sum of three terms, two of which correspond to 
the (negative values of) self-energy integrals $S$ and $S'$,
and a third (interaction) integral $I$,
\begin{subequations}
\begin{align}
\label{EG}
E_G =& \; -S - S' + I \,,
\\[0.1133ex]
\label{Sint}
S =& \; G \, \int \dd^3 x \int \dd^3 y \,
\frac{{\widetilde \rho}(\vec x) \, 
{\widetilde \rho}(\vec y) }{| \vec x - \vec y |} \,,
\\[0.1133ex]
\label{Sprime_int}
S' =& \; G \, \int \dd^3 x \int \dd^3 y \,
\frac{{\widetilde \rho}'(\vec x) \, {\widetilde \rho}'(\vec y) }{| \vec x - \vec y |} \,,
\\[0.1133ex]
\label{I_int}
I =& \; 2 G \, 
\int \dd^3 x \int \dd^3 y \,
\frac{{\widetilde \rho}(\vec x) \, 
{\widetilde \rho}'(\vec y) }{| \vec x - \vec y |} \,.
\end{align}
\end{subequations}
The easiest integral to approximate in this case is 
\begin{align}
I \sim & \; 2 G \, m_n^2 \, \int \dd^3 x \int \dd^3 y \,
\frac{\delta^{(3)}(\vec x) \, \delta^{(3)}(\vec y - \vec h) }%
{| \vec x - \vec y |} 
\nonumber\\[0.1133ex]
= & \; 2 \frac{ G \, m_n^2 }{| \vec h | } \,,
\end{align}
which is the gravitational interaction energy of two neutrons,
a distance $h = |\vec h| \sim 2.5 \, {\rm cm}$ apart
(in the experiment described in Ref.~\cite{CoOvWe1975}).
Newton's gravitational constant is denoted as $G$.

In order to estimate the order-of-magnitude of the 
gravitational self-energies $S = S'$,
we need a measure of the spread of the
mass distribution $f$, which enters the 
modulus $| \vec x - \vec y |$ in the 
integrals for $S$ and $S'$ [see Eqs.~\eqref{Sint} and~\eqref{Sprime_int}]. 
We estimate the length scale of the 
mass distribution to be equal to the de Broglie wavelength 
of the neutron, which, for the experiment~\cite{CoOvWe1975},
is equal to $\lambda = 1.445\,\mbox{\AA}$.
One obtains
\begin{equation}
S = S' \sim \frac{G m_n^2}{\lambda} \,.
\end{equation}
A numerical evaluation, using experimental parameters
given in Ref.~\cite{CoOvWe1975}, leads to 
\begin{equation}
S = S' \approx 1.3 \times 10^{-54} \, {\rm J} \,,
\qquad
I \approx 1.5 \times 10^{-62} \, {\rm J} \,.
\end{equation}
We notice that the sign of the 
energy $E_G$ defined in Eq.~\eqref{EG} 
is not a priori determined by the formalism used
and depends on details of the mass distribution.
It is obtained as a negative quantity
if the formula is applied to the experimental configuration
used in Ref.~\cite{CoOvWe1975}. This problem could be remedied by 
replacing $E_G$ by its modulus $| E_G |$ in Eq.~\eqref{tC}.
Otherwise, one might argue that it is somewhat counter-intuitive 
that smaller Compton wavelengths $\lambda$ (which occur
at higher energies)
in the self-energy integrals $S$ and $S'$
(not in the interaction integral $I$)
induce a faster gravitational collapse of the 
wave function.

Finally, for the experimental configuration described in
Ref.~\cite{CoOvWe1975}, the (modulus of the) 
time $t_C$ is obtained to be of the order of 
\begin{equation}
\label{tCestimate}
| t_C | \sim 10^{18}\,{\rm s} \,,
\end{equation}
which is longer than the age of the Universe.
Hence, the Penrose conjecture as 
given in Eq.~\eqref{EG1} predicts a collapse time 
for the gravitational interference experiment~\cite{CoOvWe1975}
which is so long that the effect can 
safely be neglected in the analysis of the 
experiment. 

One should supplement 
an estimate concerning atomic spectroscopy.
Indeed, for atomic states, one 
can easily estimate that the gravitational
self-energy integrals of the 
mass distributions associated with the
atomic wave functions~\cite{BeSa1957} are of the order of
\begin{equation}
\label{EG_atomic}
E_G \sim G \, \frac{m^2}{a_0} \sim 
1.0 \times 10^{-54} \, {\rm J} \,,
\end{equation}
where $m$ is the electron mass, and $a_0$ is the Bohr 
radius. The numerical
value of the self-energy integral~\eqref{EG_atomic}
is so small that gravitationally induced collapse of the 
wave function can be safely ignored for 
high-precision spectroscopy,
and also, for the analysis of the 
gravitational shifts discussed in Secs.~\ref{sec3} and~\ref{sec4}.

%
%
\subsection{Alternative forms}
\label{appB2}

In an alternative version of the Penrose 
conjecture, Di\'{o}si~\cite{Di1987,Di1989} has conjectured that gravitationally 
induced wave function collapse
occurs over a time scale $t_C \sim \hbar/E'_G$, where
the modified gravitational self-energy $E'_G$ 
is given by the full mass distributions,
and can be written as the sum of two interaction integrals
$\calI$ and $\calI'$,
and one self-energy integral $\calS_\oplus$,
\begin{subequations}
\begin{align}
\label{EGprime}
E'_G =& \; G \, \int \dd^3 x \int \dd^3 y \,
\frac{\rho(\vec x) \,
\rho'(\vec y) }{| \vec x - \vec y |} =
\calS + \calS' + \calS_\oplus \,,
\\[0.1133ex]
\calI = & \; 
G \, \int \dd^3 x \int \dd^3 y \,
\frac{\rho_\oplus(\vec x) \,
{\widetilde \rho}(\vec y) }{| \vec x - \vec y |}  \,,
\\[0.1133ex]
\calI' = & \; G \, \int \dd^3 x \int \dd^3 y \,
\frac{\rho_\oplus(\vec x) \,
{\widetilde \rho}'(\vec y) }{| \vec x - \vec y |} \,,
\\[0.1133ex]
\label{EGprime4}
\calS_\oplus = & \; G \, \int \dd^3 x \int \dd^3 y \,
\frac{\rho_\oplus(\vec x) \,
\rho_\oplus(\vec y) }{ | \vec x - \vec y | } \,.
\end{align}
\end{subequations}
Here, we neglect the term that does not involve $\rho_\oplus(\vec x)$.
In formulating this expression,
we have again used the fact that 
the neutrons in the Colella--Overhauser--Werner 
experiment~\cite{CoOvWe1975} are particles ``bound to the Earth''.
However, a surprising observation 
can be made if we take Eq.~\eqref{EGprime} literally. 
The expression $\calS_\oplus$, given in Eq.~\eqref{EGprime4},
is the gravitational 
self-energy of the Earth, 
\begin{equation}
E_\oplus = \frac35 \, \frac{G \, M_\oplus^2}{R_\oplus} =
2.2 \times 10^{32} \, {\rm J} \,.
\end{equation}
This huge self-energy
would induce any gravitational collapse of a wave packet
separated in the gravitational field of the Earth,
on a time scale of $10^{-67} \, {\rm s}$.


A much more intuitively sensible expression is obtained
if, instead of the product of the two mass distributions,
we use in the self-energy integral in Eq.~\eqref{EGprime} 
the difference of the two mass distributions,
$\delta\rho(\vec y) = \rho'(\vec y) - \rho(\vec y)$.
Let us therefore consider the renormalized integral
\begin{align}
\label{penrose_ren}
E''_G =& \; G \, \int \dd^3 x \int \dd^3 y \,
\frac{\rho(\vec x) \,
[\rho'(\vec y) - \rho(\vec y)]}{| \vec x - \vec y |} 
\equiv \calT \,,
\nonumber\\[0.1133ex]
\calT \approx & \;
G \, \int \dd^3 x \int \dd^3 y \,
\frac{\rho_\oplus(\vec x) \,
[{\widetilde \rho}'(\vec y) - 
{\widetilde \rho}(\vec y)]}{| \vec x - \vec y |}
\nonumber\\[0.1133ex]
= & \; m_n \, g \, h \,,
\end{align}
which is just the gravitational energy difference 
of the two wave packet contributions into which the 
neutron beam is being split in the Colella--Overhauser--Werner
experiments.

A numerical evaluation, 
with $h = \sin(22.1^\circ) \, \times 2.5 \, {\rm cm}$,
adapted to the experiment~\cite{CoOvWe1975},
leads to a value of $t_c \approx 6.8 \times 10^{-7} \, {\rm s}$
for the wave function collapse time,
if formula~\eqref{penrose_ren} is used.
This result has to be compared to the flight time of the 
neutrons in the interferometric apparatus.
Using the de Broglie relation with a neutron wavelength
$\lambda = 1.445\,\mbox{\AA}$, 
one can convert the neutron momentum 
$|\vec p_n| = m \, |\vec v_n| = h/\lambda$
into a classical velocity $|\vec v_n|$,
and, for interferometer arms of a length of 
around $2.5 \, {\rm cm}$,
to a flight time of about $t_F \approx 9.1 \times 10^{-6} \, {\rm s}$.
Because $t_C \sim t_F$,
the observation of interference fringes 
in the Colella--Overhauser--Werner
experiments~\cite{CoOvWe1975} pressures the parameters of the 
modified self-energy integral $E''_G$.
If gravitationally induced wave function 
collapse were to occur, then we would see a smearing of the 
fringes. According to Ref.~\cite{Bo2018Priv}, it would 
easily be possible to increase the arm length 
of the gravitational interferometer,
to test the renormalized form~\eqref{penrose_ren}
of the conjecture.

%
%
\subsection{Brief summary}
\label{appB3}

For systems of practical interest,
such as atomic and molecular bound states,
the original form of the Penrose 
conjecture~\cite{Pe1996penrose,Pe1998penrose,Pe2014penrose}, 
given in Eq.~\eqref{EG1}, 
predicts very long 
collapse times for quantum-mechanical wave functions,
due to gravitational effects. These are typically
long even when compared to the age of the Universe
[see Eq.~\eqref{tCestimate}].
Under reasonable assumptions, 
the collapse of the wave function can thus 
be neglected in the discussion of gravitational 
shifts or line broadenings involving 
quantum mechanical energy levels in 
bound systems.
Notably, the collapse time, when converted to 
frequency units, is smaller than the gravitational shifts
of energy levels which could lead to a
quantum limitation of the EEP.

By contrast,
the alternative form of the Penrose conjecture
proposed by Di\'{o}si~\cite{Di1987,Di1989}
[see Eq.~\eqref{EGprime}] 
fails basic consistency considerations
in regard to the 
Colella--Overhauser--Werner~\cite{OvCo1974,CoOvWe1975,BoWr1983,BoWr1984}
experiment, where a neutron wave packet is being
split in a gravitational field.
Indeed, if the conjecture were to hold in 
the form proposed by Di\'{o}si~\cite{Di1987,Di1989},
then collapse times would be so short that 
the interference fringes in the 
Colella--Overhauser--Werner experiment~\cite{OvCo1974,CoOvWe1975,BoWr1983,BoWr1984}
would disappear.

An interesting incentive for further study might 
be given by the renormalized 
form~\eqref{penrose_ren} of the Penrose conjecture,
which is proposed here.
The observation of interference fringes in the
Colella--Overhauser--Werner
experiment~\cite{OvCo1974,CoOvWe1975,BoWr1983,BoWr1984}
pressures the renormalized form of the Penrose conjecture.
However, it leads to predictions which could be tested in a 
modified form of the Colella--Overhauser--Werner 
experiment~\cite{OvCo1974,CoOvWe1975,BoWr1983,BoWr1984},
with a larger arm length for the gravitational 
interferometer. This proposal could lead to 
interesting future studies.

%
%
\section{Space--Time Noncommutativity}
\label{appC}

%
%
\subsection{Theoretical foundations}
\label{appC1}

We shall attempt to compare the parametric estimates for 
the limitation of the Einstein equivalence principle,
due to quantum effects (see Sec.~\ref{sec3}), to the 
effects that would otherwise be induced by 
space-time noncommutativity~\cite{SeWi1999,ChSJTu2001,MaSJ2001}.
The essence of the noncommutative geometry is 
to promote space-time coordinates 
to operators, which fulfill the 
commutation relations [see Eq.~(1.1) of Ref.~\cite{SeWi1999}],
\begin{equation}
[ \hat x_\mu, \hat x_\nu ] = \ii \, \theta_{\mu\nu} \,.
\end{equation}
The energy scale of space-time noncommutativity
is the upper limit for the applicability of ordinary 
quantum field theory.
Hence, it is crucial to compare the magnitude of the 
effects induced by space-time noncommutativity
to any conceivable limitations of the Einstein equivalence
principle. 

In general, one assumes that the parameters 
$\theta_{\mu\nu}$ are related to the mass scale $\Lambda_{\rm NC}$
of noncommutativity as in
\begin{equation}
\theta_{\mu\nu} \sim \frac{\hbar^2}{\Lambda_{\rm NC}^2\, c^2} \,,
\end{equation}
where we use full SI mksA units.
The original idea of Ref.~\cite{SeWi1999}
was to conjecture that $\Lambda_{\rm NC} \, c^2$ should 
be commensurable with the Planck energy, 
i.e., that its associated reduced Compton 
wavelength is equal to the Planck length~$\ell_P$,
\begin{equation}
\label{planck}
\frac{\hbar}{\Lambda_{\rm NC} \, c} = \ell_P \,.
\end{equation}
We recall that the Planck length $\ell_P$ is given by
\begin{equation}
\ell_P = \sqrt{\frac{\hbar G}{c^3}} = 1.616 \times 10^{-35} \, {\rm m} \,.
\end{equation}
So, according to Ref.~\cite{SeWi1999},
$\Lambda_{\rm NC} \, c^2$ should assume a numerical
value of the order of the Planck energy $E_p$,
i.e., the Planck mass $m_P$ multiplied by $c^2$,
\begin{equation}
\label{assumption}
\Lambda_{\rm NC} \, c^2 \sim m_P \, c^2 = 
E_p = 1.22 \times 10^{28} \, {\rm eV} \,.
\end{equation}

The authors of Ref.~\cite{ChSJTu2001}
go a different route and use Lamb shift data
in order to derive a lower bound on $\Lambda_{\rm NC}$.
To this end, they
define a vector $\vec \theta$ by the relation
$\theta^i = \epsilon^{ijk} \, \theta^{jk}$,
and assume that upon a suitable 
rotation of the coordinate system, 
they can set $\theta^3 = \theta$,
where $\theta$ is a dimensionless scalar
parameter.

According to Eq.~(3.2) of Ref.~\cite{ChSJTu2001},
the relative energy change $\delta E$, due to 
space-time noncommutativity, of a hydrogen transition
energy $E$ which involves 
a transition with a change of the principle quantum number,
is of order 
\begin{equation}
\label{deltaEbyE}
\frac{\delta E}{E} =
\alpha^2 \frac{m^2}{\Lambda_{\rm NC}^2}  \,.
\end{equation}
In Ref.~\cite{ChSJTu2001}, the authors argue that,
since theory and experiment in hydrogen agree
to a level of $10^{-13} \cdots 10^{-14}$
(see Refs.~\cite{JeKoLBMoTa2005,MoNeTa2016}),
one can derive a bound for the noncommutativity parameter $\theta$,
Specifically, 
according to the unnumbered equation following Eq.~(4.6) 
of Ref.~\cite{ChSJTu2001},
one has
\begin{equation}
\label{LambdaNCLamb}
\frac{\theta}{\lambdabar_e^2} = 
\frac{m^2}{\Lambda_{\rm NC}^2} \lesssim
10^{-7} \, \alpha \,,
\qquad
\Lambda_{\rm NC} \gtrsim 10^4 \, \frac{{\rm MeV}}{c^2} = 
10 \, \frac{{\rm GeV}}{c^2} \,.
\end{equation}
This bound is derived based on a comparison of 
Lamb shift experiments and theory.
Here, $\Lambda_{\rm NC}$ is the mass scale
of the noncommutativity of space-time.
The final numerical result
for the bound on $\Lambda_{\rm NC}$ given in the 
unnumbered equation following Eq.~(4.6) of Ref.~\cite{ChSJTu2001}
obviously contains a typographical error;
a numerical verification leads to 
values for $\Lambda_{\rm NC}$ on the order of 
$ {\rm GeV}$, not $ {\rm TeV}$.

The latest derived bounds on $\Lambda_{\rm NC}$ 
(for a summary, see Ref.~\cite{DeEtAl2017})
significantly improve over 
the paper~\cite{ChSJTu2001}.
In Sec.~IV.B of Ref.~\cite{JoChDa2015}, 
the authors arrive at a bound on the order of
\begin{equation}
\label{LambdaNC}
\Lambda_{\rm NC} \gtrsim 
\Lambda_{\rm CMBR} =
20 \, \frac{{\rm TeV}}{c^2} = 2 \times 10^4 \, 
\frac{{\rm GeV}}{c^2} \,,
\end{equation}
where the subscript CMBR denotes the
cosmic microwave background which is
measured by the Planck mission.  This improves
the bound originally derived in Ref.~\cite{ChSJTu2001} 
by more than three orders of magnitude,
and leads to a bound of
\begin{equation}
\label{dEbyEplanck}
\frac{\delta E}{E} \lesssim
\alpha^2 \frac{m^2}{\Lambda_{\rm CMBR}^2} =
3.47 \times 10^{-20} \,.
\end{equation}

Yet, on the other hand, if we assume the
order-of-magnitude estimate~\eqref{assumption}
to be valid (i.e., a scale of noncommutativity 
commensurable with the Planck scale), 
then the relative change of an atomic (hydrogen)
transition frequency is of order
[see Eq.~\eqref{deltaEbyE}]
\begin{equation}
\label{dEbyE}
\frac{\delta E}{E} \lesssim
\alpha^2 \frac{m^2}{m_P^2} =
9.32 \times 10^{-50} \,.
\end{equation}
Note that this estimate in independent of the gravitational
environment of the atom; it thus holds independently 
for the gravitational field of the Earth,
where its effect is suppressed in comparison 
to quantum limitations of the Einstein equivalence 
principle (EEP),
and also, for much more intense gravitational fields.
In the latter case, of course,
it is evidently suppressed in comparison to 
the quantum gravitational effects.

%
%
\subsection{Quantum optical experiments}
\label{appC2}

Recently~\cite{PiEtAl2012,DeEtAl2017},
a quantum optical experimental scheme has been devised whose 
aim is to dramatically improve the bounds currently 
available for $\Lambda_{\rm NC}$, with the aim 
of approaching the Planck scale.
The essential idea is to explore the noncommutative algebra
with the help of a radiation-pressure interaction of 
a micro-mechanical actuator, interacting with a 
laser beam inside a high-finesse cavity.
Specifically, the opto-mechanical effect is 
probed multiple times after the passing of the 
reference laser beam through a electro-optic modulator 
(EOM) which changes the polarization direction.
In this case, 
a sequence of four radiation-pressure interactions leads to 
an evolution operator of the form 
[see Eq.~(4) of Ref.~\cite{PiEtAl2012}]
\begin{equation}
\xi = 
\ee^{\ii \lambda n_L P_m} \,
\ee^{-\ii \lambda n_L X_m} \,
\ee^{-\ii \lambda n_L P_m} \,
\ee^{\ii \lambda n_L X_m}
\end{equation}
where $\lambda$ measures the optical path,
$n_L$ is the number of laser photons.
The dimensionless mechanical momentum and 
position operators are $P_m = p/p_0$, 
and $X_m = x/x_0$, where 
$p_0 = \sqrt{\hbar m \omega_m}$, and
$x_0 = \sqrt{\hbar/( m \omega_m )}$,
and $\omega_m$ is the mechanical resonance frequency.
Very slight deviations of the commutation relations
among the $X_m$ and $P_m$ from the canonical 
form $[X_m, P_p] = \ii$ could then be 
measured using interferometric techniques.
It is argued in Refs.~\cite{PiEtAl2012,DeEtAl2017}
that, using a high-finesse cavity with $\calF \sim 10^5$,
one could constrain $\Lambda_{\rm NC}$ to values 
approaching the Planck scale in a challenging experiment,
which would nevertheless be feasible with currently available 
technologies.

%
%
\subsection{Brief summary}
\label{appC3}

The original idea of Seiberg and Witten 
(see Ref.~\cite{SeWi1999}) was to introduce
a noncommutativity scale of the order of the 
Planck length [see Eq.~\eqref{planck}],
the underlying hypothesis being that conventional
quantum field theory breaks down
for length scales smaller than the Planck length.
We can consult Ref.~\cite{ChSJTu2001}
for an analysis of the effects of the 
noncommutativity of space-time on bound-state 
energy levels. If the original 
estimate given in Eq.~\eqref{assumption} holds,
then the effects of space-time noncommutativity 
are extremely tiny [see Eq.~\eqref{dEbyE}] 
and, notably, smaller than 
the gravitational shifts discussed in 
Secs.~\ref{sec3} and~\ref{sec4}.

One can use spectroscopic data
(see Ref.~\cite{ChSJTu2001})
or astrophysical data from the Planck 
mission (see Ref.~\cite{DeEtAl2017})
in order to formulate bounds on the 
noncommutativity scale $\Lambda_{\rm NC}$
[see Eqs.~\eqref{LambdaNCLamb} and~\eqref{LambdaNC}].
The bound~\eqref{dEbyEplanck}
is less strict than the bound~\eqref{dEbyE},
the latter being based on the 
Planck-scale hypothesis~\cite{SeWi1999}.
So, in an extreme case, the 
effects of noncommutativity 
might exceed those
discussed in Secs.~\ref{sec3} and~\ref{sec4}.

However, the original estimate given in 
Eqs.~\eqref{planck} and~\eqref{dEbyEplanck} is well motivated,
and recent proposals for ultra-precise 
quantum optical interference experiments~\cite{PiEtAl2012,DeEtAl2017}
might allow for a drastic 
improvement of the bounds for $\Lambda_{\rm NC}$,
possibly approaching the Planck scale.
In view of Eqs.~\eqref{planck} and~\eqref{dEbyEplanck},
it is indicated to assume that
the effects of noncommutativity should be 
smaller than the gravitational shifts discussed in
Secs.~\ref{sec3} and~\ref{sec4}.
Finally, we note that the quantum limitations discussed in the   
current article do not require us to consider either 
space-time quantization nor any quantization
of the gravitational interaction itself, a process which
could otherwise lead to further (very tiny) limitations of the
applicability of the equivalence principle~\cite{Gh2014}.

\end{document}